\documentstyle[12pt]{article}
\title{Parity Violation and the Nucleon-Nucleon System$^*$}
\author{W. Haeberli\\
Department of Physics\\
University of Wisconsin\\
Madison, WI  53706\\
and\\
Barry R. Holstein\\
Department of Physics and Astronomy\\
University of Massachusetts\\
Amherst, MA  01002\\}
\begin{document}
\begin{titlepage}
\maketitle
\begin{abstract}
Theoretical and experimental work seeking to understand the phenomenon of
parity violation within the nucleon-nucleon system is reviewed.
\end{abstract}
\vfill
$^*$Research supported in part by the National Science Foundation.
\end{titlepage}

\tableofcontents

\newpage

\section{Introduction}
Parity invariance has played a critical role in the evolution of our
understanding of the weak interaction.  Indeed one could argue that it was
the experiment of Wu {\it et al.}\cite{WU57} motivated by the suggestion
of Lee and Yang\cite{LEE56} that led to reexamination of the symmetry
properties of {\it all} interactions and thereby to essentially all of the
experiments discussed in this book!  Be that as it may, it is clear that
this work led in 1958 to Feynman and Gell-Mann's postulate of the $V-A$
interaction for charged currents\cite{FEYN58}, which, when combined with
Weinberg's introduction of the neutral current a decade later\cite{WEIN67},
essentially completed our picture of the weak force.  Since that time
careful experimental work has led to verification of nearly every aspect of
the proposed weak interaction structure
\begin{itemize}
\item [i)] in the leptonic sector---{\it e.g.} $\mu^-\rightarrow e^-\nu_\mu
\bar{\nu}_e, \tau^-\rightarrow e^-\nu_\tau\bar{\nu}_e$;
\item [ii)] in the $\Delta S=0,1$ semileptonic sector---{\it e.g.}
$n\rightarrow pe^-\bar{\nu}_e,\Lambda\rightarrow pe^-\bar{\nu}_e$
\item [iii)] in the $\Delta S=1$ nonleptonic sector---{\it e.g.} $\Lambda
\rightarrow p\pi^-,K^+\rightarrow\pi^+\pi^0.$
\end{itemize}
However, there is one area missing from this itemization---the $\Delta S=0$
nonleptonic interactions, {\it e.g.} $np\rightarrow np$.  Obviously there
is nothing in the identity of the particles involved to reveal the
difference between this weak interaction and the ordinary strong
$np\rightarrow np$ process.  In fact the weak NN component is dwarfed by
the much larger strong NN force but is detectable by the property of parity
violation, which it alone possesses.

On the experimental side, the first search for parity violation in the NN
interaction was carried out by Tanner\cite{TAN57} in 1957, but it was not
until 1967 that convincing evidence was presented for its existence by
Lobashov {\it et al.}\cite{LOB67}, who by using integration methods as
opposed to particle counting, was able to find a $(-6\pm 1)\times 10^{-6}$
signal among the much larger parity conserving strong background in
radiative neutron capture from ${}^{181}$Ta. That this should be the size of
a weak parity violating effect is clear from a simple scaling argument
relating the parity violating and parity conserving nucleon-nucleon
potentials $V_{NN}^{(+)}$ and $V_{NN}^{(-)}$, respectively:
\begin{equation}
{V^{(-)}_{NN}\over V^{(+)}_{NN}}\sim Gm_\pi^2\sim 10^{-7}
\end{equation}
where $G=1.01\times 10^{-5}M_N^{-2}$ is the weak coupling constant.

More than a quarter-century has
now elapsed since the Lobashov measurement and many elegant (and difficult)
experiments have been performed in this field.  Nevertheless, as we shall see,
there
remain deep and unresolved questions.  The reason for this
is that while the $\Delta S=0$ parity violating
interaction is simple at the quark level, experiments
involve, of necessity, strongly interacting hadrons, and
making a convincing connection between an experimental signal and the
fundamental Hamiltonian which it underlies has proven to be extraordinarily
difficult.  Lest one underestimate the difficulty involved, the reader is
reminded that in the related $\Delta S=1$ nonleptonic sector, the
dynamical origin of the $\Delta I={1\over 2}$ rule remains a mystery
despite three
decades of vigorous experimental activity\cite{DONO92}.
Nevertheless much has been learned in the process and it is
the purpose of this chapter to review the present situation in the field.

In doing so we are aided substantially by previous workers in this area, and
in particular by the excellent review article prepared nearly a decade ago
by Adelberger and Haxton.\cite{ADEL85}  Here we primarily summarize progress in
experiments and interpretation since that time.

\section{The Parity-Violating NN Potential}

In this section we examine the parity-nonconserving NN potential and its
relation to the underlying weak interaction from which it is derived.  Since
we will be dealing with low energy processes, we can represent the weak
interaction in terms of its local form---a point interaction of two
currents---
\begin{equation}
{\cal H}_{\rm wk}={G\over \sqrt{2}}(J_c^\dagger J_c + {1\over 2}J_n^\dagger
J_n)
\end{equation}
where, omitting contributions from the heavy (c,b,t) quarks,
\begin{eqnarray}
J_\mu^c&=&\bar{u}\gamma_\mu(1+\gamma_5)[\cos\theta_cd+\sin\theta_cs]\nonumber\\
J_\mu^n&=&\bar{u}\gamma_\mu(1+\gamma_5)u-\bar{d}\gamma_\mu(1+\gamma_5)d\nonumber\\
&-&\bar{s}\gamma_\mu(1+\gamma_5)s-4\sin^2\theta_wJ_\mu^{\rm EM}
\end{eqnarray}
are the charged and neutral weak currents respectively.
Here $\theta_c,\theta_w$ are the Cabibbo and Weinberg angles while
$J_\mu^{\rm EM}$ is the electromagnetic current\cite{BRH89}.  One set of
rigorous statements which {\it can} be made involves the isotopic spin
structure of the parity violating weak Hamiltonian, which can assume the
values 0,1,2. In particular the effective $\Delta I=2$ Hamiltonian receives
contributions only from the product of isovector charged currents---
\begin{equation}
{\cal H}_{\rm wk}^{\Delta I=2}\sim J_c^{I=1}J_c^{I=1}.
\end{equation}
On the other hand the effective $\Delta I=1$ Hamiltonian arise
s from both charged and
neutral currents---
\begin{equation}
{\cal H}_{\rm wk}^{\Delta I=1}\sim J_c^{I={1\over 2}}J_c^{I={1\over 2}}
+J_n^{I=0}J_n^{I=1}.
\end{equation}
Since $J_c^{I={1\over 2}}\propto \sin\theta_c$ and
$\sin^2\theta_c\sim 1/25<<1$, however, we expect that the primary
contribution comes from the product of isoscalar and isovector neutral
currents.  Finally, the effective $\Delta I=0$ Hamiltonian receives significant
contributions from {\it both} neutral and charged currents---
\begin{equation}
{\cal H}_{\rm wk}^{\Delta I=0}\sim J_c^{I=0}J_c^{I=0}+J_n^{I=0}J_n^{I=0}
+J_n^{I=1}J_n^{I=1}.
\end{equation}
Now while such relations are easy to write down at the quark level, their
implications for the nucleon-nucleon system are much more subtle.  The
reason is that, because of the heaviness of the W,Z, the low energy weak
interaction is essentially pointlike---of zero range.  But the
nucleon-nucleon interaction has a strong repulsion at small distances so
that the probability of nucleons interacting at short range is essentially
nil---{\it i.e.}, there is virtually no {\it direct} weak NN interaction.
Rather it is
known that the ordinary (parity conserving) low energy nucleon-nucleon
interaction $V_{NN}^{(+)}$ can be represented to a high degree of precision
in terms of a sum of single ($\pi,\rho,\omega$) and multiple meson
($\pi-\pi$) exchanges\cite{NAG75}.  We would expect then that its
parity-violating counterpart $V_{NN}^{(-)}$ can be represented in like
fashion, except that now one meson-nucleon vertex is weak and parity
violating, while the other is strong and parity conserving.  Consequently,
all of the weak interaction physics is contained within the values of these
parity violating NNM coupling constants.

Because of the ``hard core" associated with the nucleon-nucleon
interaction, it is customary to include only mesons of mass less than 800
MeV or so, and our task is further simplified by use of Barton's
theorem\cite{BAR61}, which asserts that exchange of neutral and spinless
mesons between on-shell
nucleons is forbidden by CP invariance.  Therefore only $\pi^\pm,\rho$
and $\omega$ vertices need be considered and the form of the most general
parity
violating effective Hamiltonian is easily found:
\begin{eqnarray}
{\cal H}_{\rm wk}&=&{f_\pi\over 2}\bar{N}(\tau\times\pi)_3N\nonumber\\
&+&\bar{N}\left(h_\rho^0\tau\cdot\rho^\mu +h_\rho^1\rho_3^\mu
+{h_\rho^2\over 2\sqrt{6}}(3\tau_3\rho_3^\mu -\tau\cdot\rho^\mu)\right)
\gamma_\mu\gamma_5N\nonumber\\
&+&\bar{N}(h_\omega^0\omega^\mu+h_\omega^1\tau_3\omega^\mu
)\gamma_\mu\gamma_5N
-h_\rho^{'1}\bar{N}(\tau\times\rho^\mu)_3{\sigma_{\mu\nu}k^\nu\over
2M}\gamma_5N
\end{eqnarray}
We see that there are in general seven unknown weak couplings
$f_\pi,h_\rho^0, \ldots$.  However, calculations indicate that
$h_\rho^{'1}$ is quite small\cite{BGH81} and this term is generally omitted,
leaving parity violating observables to be described in terms of just six
constants.  The means by which one attempts to determine these couplings
experimentally will be described shortly.  However, before doing so we
shall examine the theoretical predictions for the size of these vertices
from the underlying weak interaction.

\subsection{Theoretical Calculation of Weak Coupling Constants}

One of the first estimates of the weak parity violating vertex constants
was provided in the early 1960's.  F.C. Michel\cite{MICH64} estimated the
charged current couplings to vector mesons using the so-called
factorization approximation, which replaces a sum over a complete set of
intermediate states by only the vacuum state contribution---
\begin{eqnarray}
<\rho^+ n|{\cal H}_{\rm wk}^c|p>&=&{G\over \sqrt{2}}\cos^2\theta_c
<\rho^+ n|V^\mu_+A_\mu^-|p>\nonumber\\
&\approx&{G\over \sqrt{2}}\cos^2\theta_c<\rho^+|V^\mu_+|0><n|A_\mu^-|p>
\end{eqnarray}
The justification for this approximation is basically that it is possible
and easy to calculate.
There is no reason to believe that it provides anything other than an order of
magnitude estimate.

The next major theoretical development occurred in 1970 with the realization
that the charged current contribution to pion production could be written
using SU(3) symmetry in terms of {\it experimental} parity violating hyperon
decay amplitudes\cite{EQNC2}
\begin{equation}
<\pi^+ n|{\cal H}_{\rm wk}^c|p>=-\sqrt{2\over 3}\tan\theta_c
 (2<\pi^-p|{\cal H}_{\rm wk}|\Lambda^0>-<\pi^-\Lambda^0|
{\cal H}_{\rm wk}|\Xi^->).
\end{equation}
Unfortunately, this result is not as convincing as it appears, since it
involves a substantial cancellation between $\Lambda^0$ and $\Xi^-$ decay
amplitudes and is therefore rather sensitive to possible SU(3) breaking
effects\cite{MCK67}.

Three years later McKellar and Pick\cite{MCK73} showed how the symmetry
$SU(6)_W$ could
be applied to the $\Delta S=0$ parity violating interaction, thereby
relating pion and vector meson emission amplitudes. They determined that
the vector meson amplitudes predicted via symmetry were of opposite sign
and considerably larger than those given by factorization thereby ameliorating
an experimental sign discrepancy
 which existed at that time.  However, this
approach too was incomplete in that i) there were additional SU(6) couplings
which
were {\it not} predictable from experimental data and ii) because of its
non-$V-A$ character one could not
treat the neutral current Hamiltonian in terms of this approach.

A comprehensive calculation which included all previous results and which
enabled predictions to be made for all NNM couplings from both charged and
neutral current pieces of the weak Hamiltonian was performed in 1980 by
Desplanques, Donoghue, and Holstein (DDH)\cite{DESP80}.  Although
additional calculations have been performed during the intervening
years\cite{DUB86,FELD91}, nearly all are very similar in method and/or
yield numerical results which are qualitatively similar to those of DDH, so
we shall spend some time summarizing this work.

The basic idea of the work of DDH is use of the valence quark model, within
which the nucleon can be constructed in terms of three quark creation
operators
\begin{equation}
|
N>\sim b^\dagger_{qs}b^\dagger_{q's'}b^\dagger_{q``s"}|0>,
\end{equation}
where we imagine the spins, isospins to be combined to form components of
a spin, isospin doublet and the colors to be contracted to form a singlet.
Likewise we can construct the vector and pseudoscalar mesons via
\begin{equation}
|M>\sim b^\dagger_{qs}d^\dagger_{q's'}|0>
\end{equation}
using quark and antiquark creation operators.  The weak Hamiltonian
itself has a local current-current structure and involves four quark
fields
\begin
{equation}
{\cal H}_{\rm wk}\sim {G\over \sqrt{2}}\bar{\psi}{\cal O}\psi\bar{\psi}
{\cal O}'\psi .
\end{equation}
A generic NNM weak matrix element then is of the form
\begin{eqnarray}
<MN|{\cal H}_{\rm wk}|N>&=&{G\over \sqrt{2}}<0|(b_{qs}b_{q's'}b_{q``s"})
(b_{qs}d_{q's'})\nonumber\\
&\times&\bar{\psi}{\cal O}\psi\bar{\psi}{\cal O}'\psi(b^\dagger_{qs}
b^\dagger_{q's'}b^\dagger_{q``s"})|0>\times R
\end{eqnarray}
where $R$ represents a complicated radial integral.  The vacuum expectation
value is tedious to
calculate but doable.  Thus one finds
\begin{equation}
<MN|{\cal H}_{\rm wk}|N>\sim \mbox{known ``geometrical" factor}\times R
\end{equation}
which is in the form of a Wigner-Eckart theorem, where the known ``geometrical
factor" is a Clebsch-Gordan coefficient and $R$ represents a reduced matrix
element, which is identical for all such transitions and may be determined
empirically by comparing one such amplitude with its experimental value.
In fact when this procedure is followed for the simple charged current
Hamiltonian the $SU(6)_W$ results of McKellar and Pick are exactly reproduced.
However, within the quark model based procedure one can treat the neutral
current matrix elements on an equal footing.  Also, since the $\pi$- and
$\rho$-meson masses are so different it is essential to include $SU(6)$
breaking effects, and the quark model offers a means of doing this.

While details can be found in ref. 17, the results can be summarized in
terms of three different types of reduced matrix elements as shown  in
Fig.\ 1.  Figure 1a represents the factorization diagrams with the vector or
pseudoscalar meson connecting to the vacuum through either the $V$ or $A$
current respectively, multiplied by the nucleon-nucleon matrix element of the
$A$ or $V$ current.  The remaining two diagrams are of a different character
and correspond to more complicated baryonic intermediate states.  Figure
1b can be shown to correspond to the SU(3) sum rule of Eq.\ (10).  Note that
since the hyperon decay amplitudes are themselves proportional to $\cos\theta_c
\sin\theta_c$ the charged current Hamiltonian contribution to pion emission
is proportional to $\sin^2\theta_c$ and is strongly suppressed.
However, this is {\it not} the case for the corresponding
neutral current contribution, which is of ${\cal O}(1)$ and consequently
dominates
the pion emission amplitude.  Finally, Fig.\ 1c represents the new contribution
to the vector/pseudoscalar emission identified by McKellar and Pick.

\begin{figure}
\vspace{3in}
\caption{Quark model diagrams for parity violating NNM vertices.}
\end{figure}

Despite the understanding gained by connecting the quark model and symmetry
based calculations, DDH emphasized that there remain major difficulties in
attempts to provide reliable numerical estimates for these weak parity
violating
couplings.  These include uncertainty in
\begin{itemize}
\item [i)] the (large) S-P factorization term due to its
dependence on the absolute size of the current u,d quark masses;
\item [ii)] enhancement factors associated with the renormalization group
treatment of the effective weak Hamiltonian;
\item[iii)] use of a relativistic vs. a nonrelativistic quark model;
\item [iv)] the size of the sum rule contribution to pion emission
due to SU(3) breaking;
\item [v)] the size of the vector meson vs. pion emission amplitudes due
to SU(6) breaking effects;
\item [vi)] {\it etc.}
\end{itemize}
Because of all of these unknowns DDH presented their results not as a single
number but rather in terms of a {\it range} inside of which it
 was extremely
likely that a given parameter would be found.  In addition they presented a
single number called the ``best value" but this is described
simply as an educated guess in view of all the uncertainties outlined above.
The results of this process are summarized in Table 1.

\begin{table}
\caption{Weak NNM couplings as calculated in refs. 17-19.  All
numbers are quoted in units of the ``sum rule" value $3.8\times 10^{-8}$.}
\begin{center}
\begin{tabular}{|c|c|c|c|c|}
\hline
\quad   & DDH\cite{DESP80} & DDH\cite{DESP80} & ref. 18 & ref. 19\\
Coupling & Reasonable Range & ``Best" Value & DZ & FCDH\\ \hline
$f_\pi$ & $0\rightarrow 30$ &12&3&7\\
$h_\rho^0$& $30\rightarrow -81$&-30&-22&-10\\
$h_\rho^1$& $-1\rightarrow 0$& -0.5&+1&-1\\
$h_\rho^2$& $-20\rightarrow -29$&-25&-18&-18\\
$h_\omega^0$&$15\rightarrow -27$&-5&-10&-13\\
$h_\omega^1$&$-5\rightarrow -2$&-3&-6&-6\\ \hline
\end{tabular}
\end{center}
\end{table}

\subsection{Parity Violating Nucleon-nucleon Potential}

Before we can make contact with experimental results it is necessary to
convert the NNM couplings generated above into an effective parity
violating nucleon-nucleon potential.  Inserting the strong couplings,
defined via
\begin{eqnarray}
{\cal H}_{\rm st}&=&ig_{\pi NN}\bar{N}\gamma_5\tau\cdot\pi N
+g_\rho\bar{N}\left(\gamma_\mu+i{\mu_V\over 2M}\sigma_{\mu\nu}k^\nu\right)
\tau\cdot\rho^\mu N\nonumber\\
&+&g_\omega\bar{N}\left(\gamma_\mu+i{\mu_S\over 2M}\sigma_{\mu\nu}k^\nu
\right)\omega^\mu N
\end{eqnarray}
into the meson exchange diagrams shown in Fig.\ 2

\begin{figure}
\vspace{6cm}
\caption{Parity violating NN potential generated by meson exchange.}
\end{figure}

\noindent and taking the Fourier
transform one finds the effective nucleon-nucleon potential
\begin{eqnarray}
V^{\rm PNC}&=&i{f_\pi g_{\pi NN}\over \sqrt{2}}\left({\tau_1\times\tau_2\over
2}
\right)_3(\sigma_1+\sigma_2)\cdot\left[{{\bf p}_1-{\bf p}_2\over 2M},f_\pi (r)
\right]\nonumber\\
&-&g_\rho\left(h_\rho^0\tau_1\cdot\tau_2+h_\rho^1\left({\tau_1+\tau_2\over 2}
\right)_3+h_\rho^2{(3\tau_1^
3\tau_2^3-\tau_1\cdot\tau_2)\over 2\sqrt{6}}\right)
\nonumber\\
&\times&\left((\sigma_1-\sigma_2)\cdot\left\{{{\bf p}_1-{\bf p}_2\over 2M},
f_\rho(r)\right\}+i(1+\chi_V)\sigma_1\times\sigma_2\cdot\left[{{\bf p}_1-
{\bf p}_2\over 2M},f_\rho (r)\right]\right)\nonumber\\
&-&g_\omega\left(h_\omega^0+h_\omega^1\left({\tau_1+\tau_2\over 2}\right)_3
\right)\nonumber\\
&\times&\left((\sigma_1-\sigma_2)\cdot\left\{{{\bf p}_1-{\bf p}_2\over 2M},
f_\omega (r)\right\}+i(1+\chi_S)\sigma_1\times\sigma_2\cdot\left[{{\bf p
}_1
-{\bf p}_2\over 2M},f_\omega(r)\right]\right)\nonumber\\
&-&(g_\omega h_\omega^1-g_\rho h_\rho^1)\left({\tau_1-\tau_2\over 2}\right)_3
(\sigma_1+\sigma_2)\cdot\left\{{{\bf p}_1-{\bf p}_2\over 2M},f_\rho(r)\right\}
\nonumber\\
&-&g_\rho h_\rho^{1'}i\left({\tau_1\times\tau_2\over 2}\right)_3
(\sigma_1+\sigma_2)\cdot\left[{{\bf p}_1-{\bf p}_2\over 2M},f_\rho(r)\right]
\end{eqnarray}
where $f_V(r)=\exp (-m_Vr)/4\pi r$ is the usual Yukawa form.

Armed now with the form of the parity violating nucleon-nucleon potential
one can attempt to calculate the size of parity-violating observables
which might be expected in a given experiment.  However, before doing so
it is useful to examine the general types of experimental signals of
parity violation which one might look for.

\section{Experimental Signals of Parity Violation}

\subsection{Observables}

Parity refers to the behavior of a system under spatial inversion, that is
under the mathematical transformation ${\bf r}\rightarrow -{\bf r}$.  Under
spatial inversion momentum, being proportional to velocity,
also changes sign---${\bf p}\rightarrow{\bf -p}$---but {\it angular} momentum,
being an axial vector, does not---${\bf L}={\bf r}\times{\bf p}
\rightarrow -{\bf r}\times -{\bf p}=+{\bf L}$.  Likewise spin
must transform into itself under
a spatial inversion.  Thus one generally looks for a parity violating signal
by examining a correlation which is odd under spatial inversion, such as
photon circular polarization, which has the form $\sigma\cdot{\bf p}$.

{\it a) $P
_\gamma$-circular polarization in $\gamma$-decay:} That
the presence of non-zero circular polarization is a signal of parity
violation can be seen within the context of a simple example.
Consider a transition involving emission of electric and magnetic dipole
radiation, for which the relevant operators have the form
\begin{eqnarray}
E1&:& \hat{\epsilon}_\gamma\cdot {\bf p}\nonumber\\
M1&:& i\hat{\epsilon}_\gamma\times{\bf q}\cdot{\bf L}.
\end{eqnarray}
Circular polarization involves a linear combination of
 polarization states
orthogonal to the photon momentum and $90^\circ$ out of phase
\begin{equation}
\hat{\bf q}_\gamma=\hat{z}, \quad \hat{\epsilon}_{R,L}=\sqrt{1\over 2}
(\hat{x}\pm i\hat{y}).
\end{equation}
As both ${\bf p}$ and ${\bf L}$ are tensors of rank one, the Wigner-Eckart
theorem guarantees that the $E1,M1$ amplitudes are proportional
\begin{equation}
<f|{\cal O}_{E1}|i>\propto<f|{\cal O}_{M1}|i>.
\end{equation}
Finally, since $\hat{\epsilon}_\gamma , \hat{\epsilon}_\gamma\times\hat{\bf q}
,\hat{\bf q}$
are mutually orthogonal, we see that the simultaneous presence of both electric
and magnetic dipole transitions {\it must} lead to circular polarization.
However, since ${\bf p}$ is a polar vector while ${\bf L}$ is an axial
vector the selection rules are different
\begin{eqnarray}
E1&:& \Delta J=0,\pm 1;\quad P_iP_f=-1\nonumber\\
M1&:& \Delta J=0,\pm 1;\quad P_iP_f=+1
\end{eqnarray}
so that clearly a violation of parity invariance is required for the
existence of circular polarization.

While nonzero circular polarization is then a clear indication
of parity noninvariance, detection of such a signal is made difficult by the
fact that there exist no efficient circular polarization analyzers.  All such
polarimeters are based on the spin dependence of Compton scattering by
polarized electrons in magnetized iron.  However, even at saturation only
$2/26\sim 8\%$ of the Fe electrons are polarized so this represents an upper
bound for the analyzing power of such a polarimeter.  In fact, typical values
for actual instruments are typically 4\% or less.

{\it b) A$_\gamma$-asymmetry in $\gamma$-decay:}
Because of this limitation, many experiments have instead chosen to polarize
the parent nucleus and to look for the
existence of a decay asymmetry of the emitted photon with respect to
the polarization direction---{\it i.e.} a correlation $<{\bf J}>\cdot
{\bf q}_\gamma$.  The difficulty in this case is to provide a large,
reversible degree of polarization for the decaying nucleus.

{\it c) A$_z$-longitudinal analyzing power:}
A third parity violating observable is the longitudinal analyzing power of
reactions involving polarized nucleons---
\begin{equation}
A_Z={1\over P_Z}\left({\sigma_+-\sigma_-\over \sigma_++\sigma_-}\right)
\end{equation}
where $P_Z$ is the longitudinal polarization and $\sigma_\pm$ are the
cross sections for right and left handed helicity nucleons
respectively---{\it i.e.} a correlation $<{\bf J}\cdot{\bf p}>$.
Such measurements are accomplished by rapidly switching the beam helicity.
A related, but independent, observable is the analyzing
power $A_x$ defined in analogy to Eq.\ (21). This quantity is measured with
beam polarization transverse to the beam momentum, but in the scattering
plane.

{\it d) Neutron spin rotation:}  Propagation of a beam of cold neutrons through
a homogeneous sample can be described by an index of refraction, which depends
on
the forward scattering amplitude of the atoms. Inclusion of the weak
interaction
adds (coherently) a small parity-nonconserving component to the scattering
amplitude, which causes the two neutron helicity states to accumulate different
phases in passage through the medium\cite{MICH64,STOD82}.  As a consequence, a
neutron spin initially transverse (y) to it's momentum (z) undergoes a spin
rotation in the transverse (xy) plane proportional to the thickness of material
traversed, and thus acquires a x-component of polarization. The experimental
arrangement [see ref.\cite{FORT80} and Fig. 7] makes use of a sample placed
between a neutron
polarizer
 (y) and a neutron analyzer (x) at right angle to one another, with the
sample placed in between. The sample is placed alternatively before and after a
$180^\circ$ spin rotator, which reverses the x-component of neutron
polarization. In this way
the method doubles the size of the spin rotation signal and avoids many of the
instrumental problems which would have to be dealt with if a comparison were
made of counts with sample in place and sample removed.

{\it e) Parity-forbidden decay width:}
Finally, a
fourth type of experiment involves the detection of a process
whose very existence would be forbidden were parity to be conserved.  An
example is the $\alpha$ decay
\begin{equation}
{}^{16}{\rm O}(2^-)\rightarrow{}^{12}{\rm C}(0^+)+\alpha .
\end{equation}
While the detection of such a signal is a clear indication of parity
noninvariance, unlike any of the effects described above, which are
interference experiments and consequently depend on the weak matrix element
to the {\it first} power, the observable here is a rate and is therefore
{\it second} order in the parity violating weak matrix element.  The
size of the signal is then very small (B.R.$\sim 10^{-10}$ for the case
above) and must be picked out from a much larger parity conserving
background.

No matter which type of experiment one chooses, the very small magnitude of
the expected parity violating signal at the weak level involves
considerable challenge particular for the NN interaction itself where the
effects are of the order $10^{-7}$. In addition the number of feasible NN
experiments is not sufficient to
determine the separate weak NN couplings listed in Table 1. Thus many of
the experiments listed below involve studies of parity violating effects in
complex nuclei.

\subsection{Experimental Systems}

In selecting systems by which to study the phenomenon of nuclear parity
violation, one has a number of choices.  Certainly the cleanest from a
theoretical point of view is the NN system.  Indeed experimental phase shifts
are known up to hundreds of MeV and beam/target systems are readily available.
However, one pays a high price in that the expected signal is in the canonical
$10^{-7}$ range.  Thus such experiments are notoriously sensitive to tiny
systematic effects.  In fact for the np system there still exists no compelling
experimental signal.

A second arena is that of few body nuclei, {\it e.g.} p-d, p-$\alpha$
scattering and
n-d radiative capture.  In this case use of Faddeev and other methods provides
a relatively believable theoretical base.  However, it is by no means as clean
as that for the NN processes, and one still is faced with generally tiny
experimental effects, which require heroic experimental efforts.

The use of p-shell and heavier
nuclei in the study of nuclear parity violation is an alternative route, but
it has both
positive and negative implications.  On the plus side, the nuclear
environment offers an enormous assortment of various spin-parity states
which can in principle be exploited.  Also, one can in some cases
use the nucleus as an amplifier, in order to yield parity
nonconserving signals much larger than the generic $10^{-7}$
estimated above.   However, interpretation of such
experiments in terms of fundamental weak interaction parameters requires
knowledge of the nuclear wavefunctions at a level considerably more
precise than needed for the understanding of more traditional (and parity
conserving) nuclear measurements.

An excellent example of the large enhancement that sometimes occurs in
complex nuclei is provided
by the measurement of the photon asymmetry in the decay of
$8^-$ isomer of ${}^{180}$Hf, which yields a 2\% effect\cite{KRANE71}
\begin{equation}
A_\gamma = -(1.66\pm0.18)\times 10^{-2}.
\end{equation}
An even larger signal is seen in low energy neutron scattering from
${}^{139}$La, where the longitudinal asymmetry has been measured to
be\cite{YUA91}
\begin{equation}
A_L=(9.55\pm 0.35)\times 10^{-2}.
\end{equation}
In order to see how such large effects can come about, consider a nucleus
having states with identical spins but opposite
parity---say $J^+,J^-$---which are very close to one another in energy.  Now
although we have labelled such states by their spin and parity, in reality
neither state is a true eigenstate of parity, because of the presence of the
weak
interaction.  (Spin, of course, {\it is} a good quantum number because
angular momentum is exactly conserved.)  We can calculate the mixing of these
presumed close-by levels using first order perturbation theory, yielding
\begin{eqnarray}
|\psi_{J^+}>&\simeq& |\phi_{J^+}>+{|\phi_{J^-}><\phi_{J^-}|{\cal H}_{\rm wk}
|\phi_{J^+}>\over E_+-E_-}\nonumber\\
&=&|\phi_{J^+}>+\epsilon|\phi_{J^-}>\nonumber\\
|\psi_{J^-}>&\simeq&|\phi_{J^-}>+{|\phi_{J^+}><\phi_{J^+}|{\cal H}_{\rm wk}
|\phi_{J^-}>\over E_--E_+}\nonumber\\
&=&|\phi_{J^-}>-\epsilon|\phi_{J^+}>
\end{eqnarray}
in an obvious notation.  Note that we have truncated the sum over {\it all}
intermediate states down to a single state by the assumption that the two
states being considered here are nearly
 degenerate.  We can estimate the size
of the mixing parameter $\epsilon$ by scaling to a typical nuclear level
splitting, of the order of an MeV or so.  Since this splitting
is associated with the strong interaction we estimate
\begin{equation}
<\phi_{J^-}|{\cal H}_{\rm wk}|\phi_{J^+}>\sim {<{\cal H}_{\rm wk}>\over
<{\cal H}_{\rm st}>}\times 1MeV\sim 1 eV.
\end{equation}
For a pair of levels with a typical---MeV---spacing, we then have
\begin{equation}
\epsilon ={<\phi_{J^-}|{\cal H}_{\rm wk}|\phi_{J^+}
>\over  E_+-E^-}
\sim {1eV\over 1MeV}\sim 10^{-6}
\end{equation}
as expected.  However, the mixing can be substantially enhanced by selecting
two levels which are nearly identical in energy.  Thus, for example, for
two states which are separated by say 100 eV one might expect an effect of
the size
\begin{equation}
\mbox {Parity Violating Effect}\sim 10^{-6}\times {1MeV\over |E_+-E_-|}\sim
10^{-2}
\end{equation}
and the situation of ${}^{139}$La falls into this category, involving a
narrow p-wave state embedded in a host of nearby s-wave resonances.
The case of ${}^{180}$Hf involves
a 501 keV gamma ray, however, and reveals
an alternative means by which nuclear enhancement can arise.  Since the
transition connects $8^-$ and $8^+$ levels, the transition would be expected
to be predominantly electric dipole, with a small magnetic dipole component
generated by the presence of parity mixing, and the resultant asymmetry
would be of order
\begin{equation}
A_\gamma\sim 2\epsilon {<M1>\over <E1>}.
\end{equation}
However, for ${}^{180}$Hf the E1 transition is highly retarded, having
$\Delta$K=8
in the Nilssen rotational model, and this selection rule violation accounts for
the very large signal.

Despite the obvious experimental advantages to having 1\% signals to deal with
rather than the generic $10^{-6}$ effects found in direct NN experiments, the
use of complex nuclei does not permit rigorous extraction of the size of weak
effects because of the lack of believable nuclear wavefunctions for such heavy
nuclei. However, as we shall see below (Sect.\ 7) sufficiently good wave
functions have been established for a number of s-d and p shell nuclei.  In
addition, for heavy nuclei information has been extracted by use of
statistical arguments (Sect.\ 8).

\section{Proton-Proton Interaction}

The simplest system wherein the weak parity violating interaction can be
studied
consists of a pair of nucleons. Since experimental studies of the two-neutron
interaction are out of the question for obvious reasons, that leaves either the
pp or the pn system, which we shall examine in this and in the
following section.

The parity violating pp interaction has been studied by a number of groups by
measuring the analyzing power $A_z$ for longitudinally polarized protons. In
isospin space, two protons form an isotriplet and therefore the parity
nonconserving interaction in this case will involve all the isospin components
-
$\Delta I=0,1,2.$

Depending on proton energy, measurements on the pp system use one  of the
arrangements shown schematically in Fig.\ 3. At high energies, the helicity
dependence $A_z^{tot}$ of the total cross section is deduced from the change in
transmission through the sample when the spin direction of the incoming beam is
reversed, the transmission being measured by the ratio of the beam
intensity before and after the sample. At lower energies (E$<$50 MeV) the
transmission method is not useful because the large proton energy loss in the
sample
limits the useable target thickness, so that the attenuation by nuclear
interactions is too small to be measured with sufficient accuracy. Instead, one
measures the intensity of scattered particles, for both beam helicities,
divided by the intensity that passed through the sample.
To improve the statistical error, and to reduce certain systematic
errors, the detector is arranged to cover all or most of the range in azimuthal
angle.

\begin{figure}
\vspace{2.5in}
\caption{Schematic arrangement for transmission (a) and scattering (b)
experiments.}
\end{figure}

\subsection{Low
energy region}

Because of the short range of the PNC interaction, below 400 MeV only low
partial waves contribute to the PNC amplitudes, namely
the $({}^1{\rm S}_0 \leftrightarrow {}^3{\rm P}_0)$ and
the J=2 transition $({}^3{\rm P}_2
\leftrightarrow {}^1{\rm D}_2)$. The two contributions add
incoherently:
\begin{equation}
A_z(E,\theta ) = A^{J=0}_z (E,\theta ) + A^{J=2}_z (E,\theta).
\end{equation}
The {\it relative} dependence on energy and angle  of each of the two terms
can be calculated from the strong interaction phase shifts\cite{SIM86,SIM88}.
The angular dependence of the J=0 contribution is isotropic, but the
J=2 contribution shows a pronounced variation with
angle\cite{SIM88,DRI89,NES88}. The energy dependence of the angle-integrated
PNC
analyzing power  $A_z^{tot}$ is shown in Fig.\ 4. The purpose of PNC
experiments is to determine the two unknown absolute normalizations (scale
factors) which multiply the $A_z^{J=0}$ and $A_z^{J=2}$, respectively.

\begin{figure}
\vspace{2.5in}
\caption{Energy dependence of the J=0 (${}^1{\rm S}_0-{}^3{\rm P}_0$) and the
J=2
(${}^3{\rm P}_2-{}^1{\rm D}_2$) PNC transition in pp scattering, calculated
from the known
strong pp phase shifts.  The sign and absolute normalization of the vertical
scale for each of the two curves must be determined experimentally.  Here the
sign and normalization (in units of $10^{-7}$) is chosen to correspond to
predictions based on the DDH ``best'' couplings.}
\end{figure}

Below about 50 MeV $A_z$ is governed by the  J=0 transition and thus is
angle-independent. Consequently, the  angular range accepted by the experiment
is chosen
to optimize statistical and instrumental uncertainties. The pioneering
experiment at Los Alamos\cite{POT74} at 15 MeV yielded $A_z = -(1.7 \pm
0.8)\times 10^{-7}$. Soon
thereafter, a group\cite{BAL80} working at SIN (Switzerland) reported a result
of $A_z = -(3.2 \pm 1.1)\times 10^{-7}$
for a proton energy of 45 MeV,
where the $A_z$ is near its maximum value. Since $A_z$ arises almost
entirely from the J=0 transition, the factor that relates $A_z$ at the two
energies is known from theory\cite{SIM88}:
\begin{equation}
A_z(45.0\ MeV) = (1.76\pm  0.01)\times A_z(15.0\ MeV).
\end{equation}

\noindent Thus the two early results are entirely consistent.  The absolute
scale depends upon the weak parity nonconserving couplings via
\begin{equation}
f({}^1{\rm S}_0-{}^1{\rm P}_0)\approx[h^{pp}_\rho g_\rho
(2+\chi_V)+h^{pp}_\omega g_\omega
(2+\chi_S)]f^T_{0^+}
\end{equation}
where
\begin{equation}
h^{pp}_\rho=h_\rho^{(0)}+h_\rho^{(1)}+{1\over \sqrt{6}}h_\rho^{(2)}\quad{\rm
and}\quad h^{pp}_\omega=h_\omega^{(0)}+h_\omega^{(1)}
\end{equation}
are combinations of parity violating parameters (note that $f_\pi$ does not
enter due to Barton's theorem) and $f^T_{0^+}$ depends upon the model of the
strong NN potential being employed.  With the DDH best values, use of the
Reid soft-core potential yields a prediction [see ref. 27 and Table 2]
\begin{equation}
A_z(45 MeV)=-1.45\times 10^{-7}
\end{equation}
while use of the Paris potential gives a value about 30\% larger.

Work with longitudinally polarized 45 MeV protons at the SIN cyclotron
continued for a decade  in attempts to eliminate or place accurate limits
on a large number of  possible systematic errors, many of which in earlier
years would have seemed  too far fetched to worry about. The final
result\cite{KIS87} of these efforts is
\begin{equation}
A_z(45.0\ MeV) = -(1.50\pm 0.22)\times 10^{-7},
\end{equation}
\noindent where the error includes  statistical  and  systematic
uncertainties as well as limits on uncertainties in the corrections for
instrumental effects.  Scattered
protons were detected in the  angular range $\theta_{lab} = 23^\circ\ {\rm
to}\ 52^\circ$. Since $A_z$ is
independent of angle, the result can be considered to represent $A_z$ in the
total cross section. However, strictly speaking the total cross section is
poorly defined because of Coulomb-nuclear interference at very forward angles,
and there is an additional  uncertainty from the possible (small) J=2
contribution. For the total {\it nuclear} PNC analyzing power at 45 MeV, the
final
result is\cite{KIS87}:
\begin{equation}
A_z^{tot}(45.0\ MeV) = -(1.57\pm 0.23)\times 10^{-7}.
\end{equation}
\noindent Agreement with the theoretical expectation (in
both magnitude and sign)
is excellent and confirms the important role of the
nonfactorization contributions to the weak vector meson exchange couplings.

Since the above represents the most accurate result on parity
violation in hadronic interactions  to date, a brief review of the experiment
is of interest. The scattering chamber is shown in Fig. 5. A longitudinally
polarized beam of 45 MeV protons is incident on a high pressure (100 bar)
hydrogen gas target, and scattered protons are detected in a
hydrogen-filled (1 bar) ionization chamber in the form of an annular cylinder
surrounding the target. The polarized  protons are produced by ionizing
polarized atoms prepared by spin separation in an atomic beam device. The
polarization of the protons is reversed by inducing  suitable
radio-frequency transitions
between hyperfine states in the neutral atoms. In this way the polarization is
reversed without the need for any change in electric and magnetic  fields seen
by the ions, which might produce a helicity-dependent change in beam
properties.
The atoms are ionized by electron bombardment inside a solenoid. The protons
are
accelerated in a cyclotron with their polarization direction transverse. The
spin is then precessed, first by a solenoid from the vertical to the horizontal
direction, and then from the horizontal transverse direction into the
longitudinal direction by a dipole magnet. For testing purposes, the
polarization can be precessed into any desired direction by choosing the proper
current in the solenoid before the dipole magnet and in a second solenoid after
the dipole magnet. Beam current on target was $3-4\mu A$ with $P= 0.83\pm
0.02$.

\begin{figure}
\vspace{2.5in}
\caption{Scattering chamber used for measurements of $A_z$ near 45 MeV.  The
drawing shows the gas target T, Faraday cup FC, graphite beam stop C, ion
chamber IC formed by foil F and collector CO.}
\end{figure}

Scattered protons are detected by measuring the current
 in the ionization
chamber, {\it i.e.} this experiment like all others at a similar level of
accuracy
uses the so-called integral counting technique introduced originally by
Lobashov\cite{LOB67}, because it is still not feasible to reach the required
accuracy by
counting individual events. The ion chamber current is integrated during 20 ms
intervals, after which the beam properties (beam intensity, beam position, beam
diameter, spatial
distribution of spurious transverse beam polarization components) are measured
during a 10 ms interval,
before the polarization is reversed. The initial polarization direction for a
group of eight such measurements is chosen at random to reduce periodic noise.
Each 60 ms measurement has a statistical error of $3.5\times 10^{-5}$, as
determined from
the variance.
The beam properties  were
determined by beam profile monitors in which narrow graphite strips were swept
across the beam. Protons scattered by the graphite strips were detected in four
detectors, to deduce the various polarization distributions $p_x(y),\
p_y(x)$ {\it etc.} In
order to gain information about dependencies not only on variation of
transverse
polarization with position (x,y) but also with angle,  two monitors in
different
locations along the beam axis are needed to correct the data. It is relatively
easy to precess the proton polarization such
that, averaged over the beam diameter, the polarization is accurately
longitudinal. The real problem is that the polarization vector for different
parts of the beam is not perfectly uniform in direction, so that the residual
polarization components $p_x,\ p_y,$ vary with position within the beam.
Particularly
dangerous is a fist moment of $p_x$ (or $p_y$) with respect to y (or x),
{\it e.g.}\ a linear
variation of $p_x$ with y. To understand the problem, look along the beam
direction
and assume that the left half of the beam has polarization up, the right half
polarization down. The regular (parity-conserving) analyzing power $A_y$ causes
particles on the left to scatter predominantly to the left, and particles on
the right to the right. When the beam polarization is reversed, the preferred
direction is correspondingly reversed and thus the ion chamber current changes
because of geometrical effects. Note that this effect does not vanish even if
scattering chamber and beam  intensity have perfect axial symmetry. The effects
can be brought under control by accurate measurements of the polarization
profile and corresponding measurements of the sensitivity of the equipment
based
on determination of the false effect for different positions and directions of
the beam with respect to the symmetry axis of the chamber.

Errors may arise from any change in beam properties which is coherent
({\it i.e.} in
step with) reversal of the beam helicity, such as small changes in beam
position
associated with helicity reversal. For reasons of symmetry, one would
expect the false effect from coherent beam motion to vanish if the beam is
exactly on the
effective center of the scattering chamber. However, a very large sensitivity
to
vertical beam motion (false parity signal of $27\times 10^{-7}$ per $\mu m$
motion) was observed
even when the beam was on the geometric axis of the chamber\cite{WH88}. The
effect was
traced to temperature gradients in the high pressure gas target caused by beam
heating. After installation of a fast blower system, which  rapidly
recirculates
the target gas, the effect of possible beam motion (measured to be less than
$0.2\mu m$) was negligible.

Another interesting question is whether there may be small changes in beam
energy when the polarization of the beam is reversed. The changes might
result from interaction of the magnetic moment of the
polarized hydrogen atoms with magnetic fields in the ion source,
but no detailed mechanism has been established.
Nevertheless, since calculations showed that already a 1 eV change in beam
energy
out of 45 MeV would cause an error in $A_z$ of $3\times 10^{-8}$,  a
possible energy modulation was investigated. The method principally made use of
reversing the
phase of the helicity by reversing the precession solenoid in the beam line.
This reverses the sign of the true PNC signal but not the sign of the possible
energy modulation signal. That energy modulations are not such a remote
possibility after all was discovered when, for other reasons, the voltage on an
electrostatic lens prior to the cyclotron was modulated. A false signal of
$100\times 10^{-7}$ was observed due to
energy modulation. The false signal could then be
used to test the rejection of the unwanted effect by solenoid reversal. This
example suggests that in experiments at the  $10^{-8}$ level of accuracy all
spurious
error sources must be investigated even if one knows of no reason why they
should be present.

For some error sources no straightforward diagnostic methods exist, so that
their investigation may require separate, auxiliary measurements which are
comparable in effort to the PNC experiment itself.
One such example is the study
of the contribution to the ion chamber current from helicity-dependent
background, such as background arising from $\beta$-decays. The concern is that
incident and scattered protons activate various parts of the scattering
chamber,
and in the process may transfer some of their polarization to the resulting
beta
emitters, which in turn contribute to the currents in the ionization chamber
and
the Faraday cup. The effort to study these effects is considerable, since many
possible reactions in different materials are involved, and each has to be
studied separately to determine if the combination of activation cross section,
polarization transfer to the beta emitter, spin relaxation times, {\it
etc.}\ are such
that a significant error might result. For discussion of other systematic
errors see. {\it e.g.}, ref.\cite{KIS87,BAL84,LANG86}.

In view of the many possible sources of error discussed in the literature, one
may well ask how one can ever be certain that some additional  error source has
not been missed. However, by now the assumptions about error suppression have
been inspected time and again in a systematic way  by several groups working on
the problems over more than two decades, so that the likelihood of an effect
that has not been thought of is  quite remote.

A good check on the correctness of an experimental result of course is obtained
from repeating the experiments by another, independent group at a different
laboratory, with different equipment, using different tests of systematic
errors. For $A_z$ in pp scattering, the group at Bonn\cite{EVER91} has reported
a new
result at 13.6 MeV, which can be compared directly to the 45 MeV result. At the
lower energy, the measured effect is expected to be smaller by a factor
$(1.85\pm 0.01)$, but the smaller magnitude of the effect is offset in part
because some of the systematic errors are less dangerous at the lower energy.
In
particular, all effects associated with the regular,
parity-allowed transverse analyzing power are significantly
reduced. The experiment used secondary-electron emission monitors to determine
the beam position, and employed feed-back devices to stabilize the beam. Their
result
\begin{equation}
A_z(13.6\ MeV) = -(0.93\pm 0.20\pm 0.05)\times 10^{-7}
\end{equation}

\noindent can be compared with the energy-corrected 45 MeV number
$A_z = -(0.81\pm 0.12)$. The results are in excellent agreement.

\subsection{Higher Energies}

A very interesting account of the  history of the Los Alamos PNC experiments at
800 MeV energy on targets of ${\rm H}_2{\rm O}$ and liquid hydrogen, and of the
5.1 GeV
experiment on ${\rm H}_2{\rm O}$ at  the Argonne ZGS has been presented in
ref.\cite{MISC88}. The total
cross section was observed by detecting the change in the fraction of beam
transmitted though the sample as the beam helicity is reversed (Fig.\ 6). The
800 MeV pp experiment used a 1m long liquid hydrogen target. Beam pulses had a
120 Hz repetition rate and an average beam current of 1 to 5 nA. Analog
signals from ion chambers (I1, I2) which measure  the beam before and after the
target are subtracted and digitized to obtain a signal that yields the
transmission $T_+$ and $T_-$.  With a transmission $T = 0.85$, the quantity
$Z = (T_+-T_-)/(T_++T_-)$ had to measured to an accuracy of $10^{-8}$ to reach
a sensitivity in
$A_z$ to $10^{-7}$. Figure 6 shows the diagnostic equipment which was used to
monitor
the beam position, intensity, size  and net transverse polarization  for every
pulse. In addition, the variation of transverse polarization across the beam
profile was determined with a scanning target and a separate four-arm
polarimeter. One advantage of the transmission method is that the sensitivity
to
the first moment of transverse polarization is smaller than for a scattering
experiment. This is an important advantage because at the higher energies the
regular pp analyzing power is large.

\begin{figure}
\vspace{3.25in}
\caption{Experimental setup used for measurements of $A_z$ at 800 MeV by the
transmission method.  The drawing is schematic and shows the 1 m long LH$_2$
target, and integrating ion chambers I1,I2.  An alalog difference (I2-I1) is
formed before digitizing the signals to reduce round-off errors.  The beam
polarization is measured by a four-arm polarimeter P1 which detects pp events,
while the polarization profile is measured by the scanning target ST and
polarimeter P2.  Integrating wire chambers W monitored beam position and size
for each pulse.  Beam position and incident angle were stabilized with signals
from split-collector ion chambers S.}
\end{figure}

The 800 MeV (1.5 GeV/c) result\cite{YUA86}:
\begin{equation}
A_z = (2.4\pm 1.1)\times 10^{-7}
\end{equation}

\noindent is of roughly the same magnitude but opposite in sign to the results
at 45 MeV. The systematic errors are small $(0.1\times 10^{-7}$), so that the
overall
uncertainty is  governed by the statistical error, which is determined in part
by the available beam current, and in part  by detector noise due to nuclear
spallation reactions in the ion-chamber surfaces.

The theoretical analysis of this measurement is much more complex than that of
its lower energy counterparts since the energy is above the pion threshold and
inelasticity effects must be taken into account.  That the result should be
positive is clear since both S-P and P-D interference terms contribute with a
positive sign above 230 MeV.  A calculation by Oka\cite{OKA81} using
experimental phase
shifts in order to unitarize Born amplitudes yields a result, using DDH best
values, which is about a factor of two above the experimental
number.  However, this calculation omitted short range correlation effects,
which tend
to reduce the size of the predicted effect considerably.  A crude estimate of
such effects made by Adelberger and Haxton\cite{ADEL85} actually reduced
the predicted
effect below the experimental number.  Subsequent work by Silbar
{\it et al.}\cite{SIL89}
attempted to model inelasticity effects by including delta degrees of
freedom and indicated an additional positive contribution of order $0.9\times
10^{-7}$. However, this was
based upon the DDH ``best" value for $f_\pi$ which as we
shall see is probably too large.  We conclude that although no definitive
calculation
exists at present, existing calculations seem to agree reasonably
well with the experimental value.

The measurement of the helicity dependence of the total cross section of 5.1
GeV
(6 GeV/c)  protons on a target of ${\rm H}_2{\rm O}$ at the Argonne
zero-gradient synchrotron
used a spectrometer to eliminate the helicity-dependent background which would
otherwise arise from hyperon decay products. The result\cite{LOCK84}
\begin{equation}
A_z = (26.5\pm 6.0\pm 3.6)\times 10^{-7}
\end{equation}

\noindent is considerably larger than is expected from most theoretical
estimates, which tend to give numbers which are positive but which are about
an order of magnitude smaller.  The discrepancy only increases if one takes
into account that the observation is for ${\rm p-H}_2{\rm O}$ rather than
p-N (on account of Glauber
corrections, see ref.\cite{LLF81}). Of course, at these energies a simple
meson-exchange potential model is no longer credible and so other
techniques---{\it e.g.} Regge theory---must be employed.  The only credible
estimate
which has thusfar been able to reproduce the ZGS measurement is a model which
involves mixing in the quark wavefunctions to negative parity excited states
via
quark-diquark interactions in the nucleon\cite{GOLD86}.  Such a model is quite
speculative and is
certainly not able to be extended to low energies in order to match onto other
calculations.  For further comments on the analysis of the 6 Gev/c result,
see refs.\cite{MISC88,GOLD86,SIM87,GOLD93,SIM93}. Certainly, a remeasurement of
asymmetry in this energy
region would be most welcome.

\subsection{Proposed and Planned Experiments}

The low-energy experiments (13.6 MeV and 45 MeV) discussed above yield
information only about  the $J=0\ ({}^1{\rm S}_0 \leftrightarrow
{}^3{\rm P}_0)$ transition.
Figure 4 shows that
starting at about 100 MeV, the $J=2$ transition $({}^3{\rm P}_2 \leftrightarrow
{}^1{\rm D}_2)$ contributes
significantly. In order to separate the two contributions, the preferred energy
is near 230 MeV, where
the $J = 0$ amplitudes cancel. Therefore the  contribution to PNC associated
with the  $J=2\
({}^3{\rm P}_2 {\leftrightarrow} {{}^1{\rm D}_2})$  amplitude can be
measured separately. This would yield an independent
determination of the $\rho$ weak coupling constant $h^{pp}_\rho$. An experiment
in this energy
range, to be carried out at TRIUMF, has been in preparation for some
time\cite{BIR88}. Two separate experiments are planned, one detecting the
helicity
dependence in transmission and one in scattering. The two experiments yield the
same information about the weak amplitudes, but they serve as an extremely
valuable test of systematic errors because the corrections are bound to have
quite different characteristics. The transmission experiment is to use a 40 cm
long liquid hydrogen target, with incident and transmitted protons detected by
ionization chambers similar to the 800 MeV experiment mentioned above. The
distribution of unwanted transverse polarization components is to be determined
by beam scanners similar to the  45 MeV experiment. A feedback system is
planned
to stabilize beam position  to 1$\mu$m.  The expected
value\cite{SIM88,DRI89} of $A_z$ is about
$0.6\times 10^{-7}$ which is to be measured to an accuracy of $0.2\times
10^{-7}$ or better. It is to be noted that in this
case the angular distribution of $A_z(\theta $) is far from isotropic.

An experiment near 230 MeV, as well as an extension to 1.5 GeV,  is also
planned  to be carried out with protons extracted from the proton storage ring
COSY at J\"ulich\cite{EVER91b}. The beam will be injected and accelerated
in the storage ring, which has provision for phase space cooling of the beam.
This is expected to  result in an extracted beam of high ion optic quality,
which in turn should reduce systematic effects associated with changes in beam
properties. The possibility to carry out experiments at much higher energies
(100 GeV) using the RHIC accelerator now under construction has been
discussed {\it e.g.} by Tannenbaum\cite{TAN93}.

It recently has been pointed out by Vigdor\cite{VIG93} that experiments with
internal targets
in storage rings may have important advantages over more conventional methods.
In particular, it is proposed to arrange precession solenoids in a storage ring
in such a fashion that only the longitudinal spin direction is stable, while
the
transverse components average to zero.

\section{Neutron-Proton Interaction}

\hspace{.5cm}{\it a) $P_\gamma$ in np-capture}:  As previously mentioned,
the first clear experimental evidence for parity violation in nuclei was
provided in a measurement by Lobashov {\it et al.}\cite{LOB67}, which
detected a nonzero
circular polarization [$P_\gamma  = (-6 \pm 1)\times 10^{-6})$] of
$\gamma$-rays from
neutron capture in $^{181}{\rm Ta}$. The experiment is known for the elegant
idea to
use a pendulum in vacuum to detect and store the repeated effects of the
small periodic signal which resulted in the $\gamma$ detector from the
reversal of the magnetic field in the magnetized iron which served as the
$\gamma$-ray polarimeter. Later, the same arrangement was used in the
first attempt to detect parity violation directly in the NN interaction.
The first result\cite{LOB72} was later
found to be contaminated with circularly polarized  bremsstrahlung caused
by  polarized electrons from beta decays of fission products in the
reactor. A new experiment\cite{KNY84}, which yielded $P_\gamma  = (1.8 \pm
1.8)\times 10^{-7}$,
was already discussed in a previous review\cite{ADEL85}. The new measurement
is consistent with the ``best value" prediction of
$0.57\times 10^{-7}$ (Table 2).  The result is
important in that it removes the earlier significant discrepancy with theory.
However, to reach an accuracy sufficient to
contribute to the determination of weak coupling constants ({\it e.g.},
$\pm 0.2\times 10^{-7}$) is probably not realistic, since the magnitude of the
experimental signal to be detected is another factor of 20 smaller on
account of the relatively small analyzing power  (0.045) of
the $\gamma$-ray polarimeter. Experimental results consistent with zero can
yield significant constraints on the determination of weak coupling
constants in those cases where two coupling constants contribute terms of
similar magnitude but opposite sign, but this is not the case for $P_\gamma$.

{\it b) Helicity Dependence in Deuteron Photodisintegration}:  As an
alternative to measuring $P_\gamma$ in np capture, one may choose to
study the inverse reaction, {\it i.e.} photodisintegration of the deuteron near
threshold  with circularly polarized photons. The photodisintegation cross
section $\sigma^+$ and $\sigma^-$  is measured with incident photons of
positive and
negative helicity to determine the parity-violating analyzing power $A_L$
defined in Eq.\ (21) where $P_L$ now refers to the photon circular
polarization. Earle {\it et al.}\cite{EAR88} accelerated  polarized electrons
produced by
photoemission from GaAs in the Electron Test Accelerator at Chalk River,
Canada to energies of 3.2 MeV or 4.1 MeV in order to produce polarized
bremsstrahlung in a water-cooled tantalum radiator. The highest energy
photons have a circular polarization equal to that of the incident
electrons, or about $P_\gamma = 0.35$.  The photons are incident on a target of
deuterated water. The photoneutrons are thermalized in the target and are
detected  via the ${}^{10}{\rm B(n},\alpha)$ reaction in boron-lined ionization
chambers. The
ionization chamber current caused by background photons was eliminated by
subtracting the current in a second set of chambers, interspersed with the
first, but without the $^{10}{\rm B}$ lining.

The effects of changes in electron beam
properties (intensity, energy, position, beam size) associated with reversal
of the beam helicity were studied in separate test experiments, and
corresponding corrections were applied to the data. It is in fact the
uncertainty of these corrections, and not statistical uncertainties, which
limit the accuracy of the data.  The final result, $A_L = (27\pm 28)\times
10^{-7}$ for
bremsstrahlung with an endpoint of 4.1 MeV, and  $A_L = (77\pm 53)\times
10^{-7}$  for an
endpoint energy of 3.2 MeV unfortunately are not accurate enough to provide
significant constraints on the weak coupling constants. A number of
improvements in the experiment were discussed by the authors,
suggest that with a major effort their method might be capable of a
sensitivity comparable with the theoretical prediction. However, the required
factor of 100 reduction in systematic and statistical errors would probably
require a group effort of at least
a decade.

\noindent {\it c) $A_\gamma$  in  np-capture}:  The quantity measured in the
above deuteron
experiments arises
from the $\Delta I=0,2$ mixing effects in the ${}^1{\rm }S_0-{}^3{\rm P}_0$ and
${}^3{\rm S}_1-{}^1{\rm P}_1$ channels.  In contrast, sensitivity to the
$\Delta I=1$ component of the
effective weak Hamiltonian
is provided by a measurement of the asymmetry in the capture of
polarized thermal neutrons.  This observable is sensitive to $\Delta I=1$
mixing effects
in the ${}^3{\rm S}_1-{}^3{\rm P}_1$ channel and thereby to
$f_\pi$.  Much evidence points
to a value of
$f_\pi$ significantly smaller than the expected weak current enhancement
predicted by DDH.  On the other hand
the measurements on  $P_\gamma$ in the decay of $^{21}{\rm Ne}$, to be
discussed
below, provide contradictory
evidence\cite{ADEL85}.   Except for negligible contributions from
$h_\rho^{(1)}$
{\it etc.},  $A_\gamma$  in
thermal np capture is directly proportional to $f_\pi$:  $A_\gamma =
-0.11\times f_\pi$. For
the DDH ``best guess" value of $f_\pi$,  $A_\gamma=  -0.5\times 10^{-7}$. ({\it
cf.} Table 2).

In view of the small expected effect, the demands on a measurement of
$A_\gamma$
are very high, but on the other hand the neutron polarization can be made
large so that one gains a large factor compared to the small analyzing
power in the $P_\gamma$  experiment. The $A_\gamma$  experiment became feasible
with  the
development of intense beams of cold polarized neutrons ($5\times 10^9$
neutrons/s
over $3\times 5 cm^{2}$)  from the high flux reactor at the
Institut Laue-Langevin
(ILL).  In the experiment described by Alberi {\it et al.}\cite{ALB88,CAV77},
capture of the neutrons takes place in a 23 liter liquid hydrogen
target, in which the hydrogen was converted to pure parahydrogen by a
catalyzer  in order to avoid depolarization of the neutrons  by scattering.
Two large tanks of liquid scintillator (400 liter volume each) detected the
2.2 MeV photons. The neutron polarization  ($P = 0.70\pm 0.07$) was reversed
about once a second by passing the neutrons through a thin current strip.
By comparing the ratio of count rates in the two detectors for the two
opposite neutron spin directions the differences in detector efficiency and
in neutron flux cancels. Systematic errors considered in the experiment
included: (i)  variation in the neutron flux with time (fluctuation
about 0.1\%) so that a small residual error remains after averaging over all
1s measurements; (ii) effect of spin flipper magnetic fields on the
detectors, compensating coils and shielding reduced these effects to
negligible proportions; (iii) displacement of the neutron beam arising from the
interaction of the magnetic moment with inhomogeneous magnetic fields; and
(iv) spurious electrical effects on the electronic circuits, such as a
shift in discriminator level when the power to the spin flipper is turned
on.  It is understood that the troubling effects are those for which a
reversal of the neutron spin has a spurious effect on the count rates
without being associated with the true parity-violation signal.
In this, as
well as other experiments at the same level of accuracy, spurious electronic
effects are avoided by making an overall change of the phase of the
polarization reversal relative to the measurement cycle. In the present
case a second spin flipper which was reversed every 27s was used for this
purpose.

The final result\cite{ALB88}, $A_\gamma = (-0.15\pm 0.47)\times 10^{-7}$, is
consistent with, but four times more accurate, than  an earlier result obtained
by the same method\cite{CAV77} [$A_\gamma
= (-0.6\pm 2.1)\times 10^{-7}$. The
new result is limited by statistical uncertainties. It is thought that,
given more running time on a suitable high flux reactor,  another order of
magnitude improvement in accuracy could be achieved (AL88). This would at
last settle the question of neutral current enhancement of the isovector
pion exchange coupling constant. For now the above result is still
consistent with the DDH best guess for $f_\pi$ (see Table 2).

{\it d) Neutron Spin Rotation}:  When transversely
polarized slow neutrons pass through
matter, parity-violating forces rotate the neutron polarization direction
about the momentum direction. Coherent rotation was proposed as a method to
detect parity violation by F.C. Michel\cite{MICH64} already in 1964.
Parity-nonconserving
neutron spin rotation was first demonstrated  experimentally in 1980 by
Forte {\it et al.}\cite{FORT80} when
transmission of cold neutrons (polarization $P_n = 0.91$)  through
$^{117}{\rm Sn}$  was shown to exhibit an unexpectedly large rotation angle
per
cm of tin of $\phi = (36.7\pm 2.7)\times 10^{-7}$ rad/cm.  The experiment
demonstrated
that neutron spin rotation is a viable tool to study parity nonconservation
in nuclei.

\begin{figure}
\vspace{2.5in}
\caption{Figure 7: Arrangement to measure PNC neutron spin rotation.}
\end{figure}

A measurement of neutron spin rotation in hydrogen would serve much the
same purpose as the above measurement of $A_\gamma$, in that both quantities
depend almost exclusively on the pion weak coupling constant. A calculation
based on the DDH ``best guess" weak coupling constants by Avishai and
Grange\cite{AVI84} predicts $\phi  = -8.84\times 10^{-9} rad/cm$,
when the Paris potential was used to describe the strong NN
interaction (see Table 2).  The Seattle group\cite{HKL} has proposed an
arrangement similar to Fig.
7, using a 25 cm thick sample of parahydrogen that is pumped back and
forth between two containers in positions 1 and 2.

\section{Few Nucleon Systems}

There exist several parity experiments on few body systems which are, strictly
speaking not NN measurements, but which are of note because they are
amenable to reasonably precise analysis.

{\it a) Polarized Thermal Neutron Capture on Deuterium.} One example
is a polarized thermal
neutron capture measurement on deuterium---$\vec{\rm n}{\rm d} \rightarrow
{\rm t}\gamma$
for which the most recent determination\cite{ALB88} has yielded $A_L= (42\pm
38)
\times 10^{-7}$ more on ref.\cite{ALB88,AVE84}.

{\it b) $A_z$ in $p-d$ and $p-\alpha$ Scattering.}
The techniques used to measure the longitudinal analyzing power in pp
scattering
have been applied to scattering of protons by helium (46 MeV\cite{LANG85b}) and
by
deuterium  (15 MeV,\cite{POT74} 45 MeV\cite{KIS89b}, 800 MeV\cite{MIS463}).  It
should be emphasized at the outset, that experiments in which
the scattered particles are detected over a certain angular range should not
be
interpreted as a measurement of the helicity dependence in the {\it total }
cross
section. Rather, there is every reason to believe that $A_z$ has a strong
dependence on angle which needs to be taken into account. This requires the
experimenters to determine not
only the angular acceptance function of the apparatus, but also the
relative contribution from inelastic channels apparatus accepts only elasticity
scattered particles or breakup products as
well.

In principle, the wave functions of the target nuclei are sufficiently well
known that the measured $A_z$ can be interpreted in terms of contributions from
weak NN coupling constants, but considerable theoretical work is required to
determine the expansion coefficients. The task is made more difficult if the
experiments include breakup channels.

{\it $p-\alpha$ scattering:}  Elastic ${\it p}-\alpha$ scattering at low
energies has attractive
theoretical and experimental features, such as simple nuclear structure and
high
breakup threshold. On the other hand, experiments on ${\rm p}-\alpha$
scattering are
even more difficult than for pp, because in this energy range the regular,
parity-allowed transverse analyzing power for ${\rm p}-\alpha$ scattering is
much larger than
for pp scattering, so that the corrections for first moments of transverse
polarization (see pp scattering above) require special attention. In fact, the
large sensitivity to transverse polarization in ${\rm p}-\alpha$ scattering was
exploited in
the  15 MeV pp  experiment by substituting He
for the H target to deduce the magnitude of the unwanted first moments of
transverse polarization in the proton beam\cite{KIB81}. At somewhat higher
energies the
situation is more favorable but still difficult.
An unpleasantly
large sensitivity to transverse polarization in a first
experiment\cite{HEN82} at
46 MeV was later reduced by an order of magnitude by redesigning the
angular acceptance function of the apparatus\cite{LANG85b}. The angular
acceptance was
chosen to take advantage of
the sign reversal of the transverse ${\rm p}-\alpha$ analyzing
power to reduce the unwanted effects, while at the same time  accepting a range
of scattering angles (primarily $\theta = 30^\circ - 60^\circ$) where the
sign of $A_z(\theta )$ does not
change. In addition, to simplify the theoretical interpretation, the wall
thickness of the target vessel was chosen such that only elastically scattered
protons have sufficient energy to penetrate the wall. The result of the
improved experiment\cite{LANG85b} is
\begin{equation}
<A_z^{{\rm p}-\alpha}(\theta ),46\ MeV>=-(3.34\pm 0.93)\times 10^{-7},
\end{equation}
where the error includes statistical and systematic errors.

Theoretical analysis here is not as simple as the corresponding
$\vec{\rm p}{\rm p}$
case wherein only knowledge of the strong phase shifts is
required.  The problem is that because of $\alpha$ particle structure
one needs not only the phase shifts but also the short distance
behavior, which in turn requires knowledge of the short range NN
correlations.  Roser and Simonius\cite{ROS85} compared the result of the above
experiment
to calculations of $A_z(\theta )$, in
which the PNC scattering amplitudes were calculated with properly
antisymmetrized optical model wave functions. The optical model takes into
account absorption from the elastic channel, but the calculation does not
include breakup channels as intermediate states in the matrix elements. The
reliability and parameter dependence of the calculations was studied in detail.
The result for $A_z$ in terms of the meson exchange coupling constants are
shown in Table 2. The short range
correlations are based on hard repulsion.   The vector meson  ($\rho$ and
$\omega$) contributions are more sensitive to short range correlations than
are the corresponding pion terms.  With a ``soft" short-range correction factor
(Jastrow factor), $A_z$
is roughly a factor two larger in magnitude $(-6\times 10^{-7})$.
 The same is seen  for pp scattering (compare Tourreil-Sprung
supersoft core with Reid soft core\cite{SIM83}).  The constraints on meson
exchange weak coupling constants provided by this experiment are very
similar to that given by the $^{19}F$ measurement discussed below.

{\it p-d scattering:}  Of the three results reported for $p-d$ scattering,
only one is a measurement of the total cross section.  The
longitudinal analyzing power in the p-d total cross section at 800 MeV
proton energy was measured at Los Alamos\cite{MIS463} by measuring the helicity
dependence of the absorption in a 1m long liquid deuterium target, using
the same equipment and methods as used for the 800 MeV pp experiment. The
largest correction [$(3.74\pm 0.37)\times 10^{-7}$] comes from the
intensity modulation associated with helicity reversal. However, the
sensitivity to these effects could be measured accurately by inserting a
grid into the $H^-$ particle beam. In this way about 10\% of the beam
particles lose their electrons so that the resulting $H^+$ ions can be
removed from the beam to change the beam intensity without changing other
beam parameters. The result of the experiment, is $A_z =(1.7\pm  0.8\pm
1.0)\times 10^{-7}$.

The two  results at lower energies (15 MeV and 43 MeV) used essentially the
same equipment and the same methods as the
corresponding pp experiments. Both experiments are based on detection of
scattered particles over a  limited angular range and thus do not measure $A_z$
in the total cross section. To complicate matters
further, the experiments do not  separate elastic scattering from break-up,
because the small binding energy of the deuteron makes it impossible to
distinguish elastic and breakup events in the integral-counting method (current
integration).  Thus a theoretical analysis would have to integrate not only
over  the
appropriate range of scattering angles
but also over the part of the breakup
phase space that is detected in the experiment, taking into account the
corresponding weight and acceptance function of the apparatus. So far,
calculations of $A_z$ have been reported only for the total elastic p-d cross
section. Thus they cannot be compared to the experimental results and
consequently no entry for p-d scattering is shown in Table 2. The
Faddeev calculations of Kloet {\it et al.}\cite{KLO83,DES79} predict for the
total cross section
$A_z^{tot}$  values
of
 $-1.87\times 10^{-7}$ at 14.4 MeV and $+1.39\times 10^{-7}$ at 40 MeV.

The 43 MeV experiment at SIN\cite{KIS89b} chose the wall thickness of the
target vessel
such that only one proton in a given breakup event reaches the detection
system. This considerably simplifies the calculation which, however, is
still very difficult. At 43 MeV, the experimental result for pd elastic
scattering and breakup protons in the laboratory angular range $24^\circ$
to $61^\circ$ is $(+0.4\pm 0.7)\times 10^{-7}$. The largest correction by
far [$(-3.25\pm 0.30)\times 10^{-7}$] is for modulation of the transverse
polarization moments. The angular acceptance function for different
Q-values of the breakup spectrum for this experiment is known, so that a
calculation for  a realistic comparison to the experiment is possible in
principle. On the other hand, for the earlier 15 MeV
measurement\cite{POT74} ($A_z = -(0.35\pm 0.85)\times 10^{-7}$)
the acceptance function has not  been specified.

{\it c) $^6Li(n,\alpha )^3H$ reaction with
polarized cold neutrons.}
Studies of PNC asymmetries in the reactions ${}^6{\rm Li(n},\alpha )^3
{\rm H}$ and
${}^{10}{\rm B(n},\alpha )^7{\rm Li}$ in which
polarized thermal neutrons are captured with large cross
sections have been suggested by Vesna {\it et al.}\cite{VES83}. The development
of high flux
beams of cold neutrons at  the VVR-M reactor at the Leningrad Institute of
Nuclear Physics has made possible a much improved determination of the helicity
dependence in the ${}^6{\rm Li(n},\alpha )^3{\rm H}$ reaction\cite{VES90}. A
multisection proportional
chamber was irradiated with  cold neutrons (average wave length 4\AA ) of
intensity
$2\times 10^{10}n/s$ and polarization 80\%.
The chamber consisted of 24 double chambers arranged along the path of the
neutron beam, with half of each double chamber detecting tritons emitted along
the direction of the neutron momentum, the other half detecting tritons in the
opposite direction. Each chamber has its own target of  ${}^6{\rm LiF}$
deposited on thin
Al foils. About 90\% of the neutron beam was absorbed in the chambers. Possible
left-right asymmetries in the chambers were reduced to the point where their
contribution to the final result is expected to be less than  $10^{-8}$. The
neutron
polarization was changed by means of an adiabatic flipper.  The measured
asymmetry coefficient  $(-0.64\pm 0.55)\times 10^{-7}$  is much smaller than
the
theoretical estimate\cite{NEST88,OKU94} based on a cluster model of
${}^6{\rm Li}$ (see Table 2).
The disagreement between the experimental value and
the theoretical
number calculated with the DDH best guess values is yet another indication
that $f_\pi$ is considerably
smaller than the ``best guess" value.

\begin{table}
\caption{Expansion coefficients for the contributions to the PNC
observables from the individual meson-exchanges and calculated observables
for three sets of coupling coefficients.
  All numbers
should be multiplied by the factor $10^{-7}$.}
\vspace{18cm}
\end{table}

\section {Isolated Parity-Mixed Doublets (Two-Level Systems)}

\subsection {Experiments}

Little benefit is gained from observations of PNC in hadronic interactions
unless the results can be interpreted to yield information about either the
weak
or the strong part  of the NN interaction, depending on whether one considers
the hadronic weak interaction (weak coupling constants) or the short range
behaviour of the strong
 interaction to be the most interesting part of the problem.
Except for the cases discussed above, in which experiments on the
nucleon-nucleon system and few-body systems have given interpretable results,
the most important source of information derives from experiments on light
nuclei, in which PNC effects result from the interference of two relatively
isolated levels of the same total angular momentum but opposite parity. As
discussed in a previous review\cite{ADEL85} the observed
effects are much magnified compare to the small effects in the NN system
provided the interfering levels are closely spaced and the members of the
parity
doublet have very different decay amplitudes. Overall, the larger magnitude of
the effects to be measured (typically $10^{-4}$ to $10^{-5}$ compared to
$10^{-7}$ in the NN
system itself) simplifies the experiments.   Regrettably, the larger effects
are
at the expense of a difficult burden in determining the nuclear structure of
the
states involved with sufficient accuracy. For the experiments, the observations
on parity-mixed doublets require different  experimental techniques: the
expected effects are large enough that sufficient statistical accuracy can be
obtained by detecting individual events (as opposed to the integral counting
techniques used for  NN and few-body experiments). This then permits sufficient
energy resolution in the detection system to isolate the levels of interest.

Until a decade ago,  studies of parity-mixed doublets in light
nuclei
(${}^{18}{\rm F},\ {}^{19}{\rm F},\ {}^{21}{\rm Ne})$ concerned primarily
gamma-decay measurements,
in particular the gamma asymmetry $A_\gamma$ in the decay of polarized
${}^{19}{\rm F}$,  and the circular polarization $P_\gamma$ of decay gamma rays
from unpolarized nuclei $({}^{18}{\rm F},\ {}^{21}{\rm Ne})$. The results of
these
experiments, which have been discussed extensively in the previous review
by Adelberger and Haxton\cite{ADEL85}, are summarized in Table 3.  Since
the transitions in these three nuclei essentially exhaust the available
pool of favorable particle-bound parity-doublets, the search for additional
parity doublets turned to particle-unbound states in light nuclei, even
though the higher excitation energy of these states tends to complicate the
nuclear structure issues.  The only new experiments on narrow, parity-mixed
doublets are measurements of the longitudinal analyzing power $A_z$ in
$({\rm p},\alpha$)-reactions, specifically ${}^{19}{\rm F(p},\alpha
){}^{16}{\rm O}$ and
${}^{13}{\rm C(p},\alpha ){}^{10}{\rm B}$. Besides $A_z$, another signal of
parity nonconservation is the  transverse analyzing power $A_x$,
{\it i.e.}, a
measurement with polarization transverse to the beam momentum but in the
scattering plane. A measurement of $A_x$ has been reported for
${}^{19}{\rm F(p},\alpha ){}^{16}{\rm O}$ (see Table 3). The general theory of
parity
mixing of elastic scattering resonances (and in particular the application
to ${}^{14}{\rm N}$) has been discussed by Adelberger, Hoodbhoy and
Brown\cite{ADEL84}.  Study of a J=2 doublet in ${}^{16}{\rm O}$ near 13
MeV excitation energy  has been been
 proposed by Bizetti and
Maurenzig\cite{BIZ80}.
Extensive calculations of the longitudinal and transverse analyzing powers
for different models of the weak and strong interactions have been reported
by Dumitrescu\cite{DUM91} and by Kniest {\it et al}\cite{KNI91}.

\begin{table}
\caption{Experimental and DDH ``best" theoretical values for parity violating
experiments in p- and s,d-shell nuclei.  }
\begin{center}
\begin{tabular}{|c|c|c|c|c|}\hline
\quad & Excited & Measured& Experiment & Theory\\
Reaction & State & Quantity& ($\times 10^{-5}$)&($\times 10^{-5}$)\\
\hline
${}^{13}{\rm C(p},\alpha){}^{14}{\rm N}$&J=$0^+$, T=1& $[A_z(35^\circ)$&$0.9\pm
0.6$\cite{ZEP94} & -2.8\cite{ADEL84}\\
\quad& 8.264 MeV & -$A_z(155^\circ)]$&\quad&\quad\\
\quad & J=$0^-$, T=1&\quad &\quad &\quad\\
\quad & 8.802 MeV & \quad &\quad &\quad\\
\hline
${}^{19}{\rm F(p},\alpha){}^{20}{\rm Ne}$& J=$1^+$, T=1&
$A_z(90^\circ)$&$150\pm
76$\cite{KNI91}&\quad\\
\quad&13.482 MeV&$A_z$&$660\pm 240$\cite{OH81}&\quad\\
\quad&J=$1^-$, T=0&$A_x$&$100\pm
100$\cite{KNI83}&\quad\\
\quad&13.462 MeV&\quad&\quad&\quad\\
\hline
${}^{18}$F& J=$0^-$, T=0& $P_\gamma$& $-70\pm 200$\cite{BARN78}& $208\pm
49$\cite{ADEL85} \\
\quad& 1.081 MeV&\quad&$-40\pm 300$\cite{BIZ80b}&\quad\\
\quad&\quad&\quad&$-100\pm 180$\cite{AHR82}&\quad\\
\quad&\quad&\quad&$17\pm 58$\cite{PAG87}&\quad\\
\quad&\quad&\quad&$27\pm 57$\cite{BINI85}&\quad\\
\quad&\quad&mean&$12\pm 38$&\quad\\
\hline
${}^{19}$F&J=${1\over 2}^-, T+{1\over 2}$& $A_\gamma$& $-8.5\pm
2.6$\cite{ADEL83} &
$-8.9\pm 1.6$\cite{ADEL85} \\
\quad& 0.110 MeV&\quad&$-6.8\pm 2.1$\cite{ELS87}&\quad\\
\quad&\quad&mean&$-7.4\pm 1.9$&\quad\\
\hline
${}^{21}$Ne&J=${1\over 2}^-, T={1\over 2}$& $P_\gamma $& $80\pm
140$\cite{SNO83} &
46\cite{ADEL85}\\
\quad&2.789 MeV&\quad&\quad&\quad\\
\hline
\end{tabular}
\end{center}
\end{table}

To illustrate the experimental methods and problems in measurements of
$A_z$ in resonance reactions, as opposed to the corresponding measurements
{\it e.g.} in pp scattering, we discuss the recent
measurement\cite{ZEP89,ZEP94} of the  analyzing power $A_z$  in the
${}^{13}{\rm C(p},\alpha ){}^{10}{\rm B}$ reaction. The experiment uses
longitudinally polarized protons near 1.16
MeV to excite  a narrow ($\Gamma$=4 keV) J=$0^+$ (T=1) state in
${}^{14}{\rm N}$ at $E_x$ =
8.624 MeV. This state interferes  with a second, much wider state ($\Gamma$=440
keV)  of opposite parity located 178 keV above the first. Therefore a small
admixture of the short-lived $0^-$ level into the long-lived $0^+$ level will
amplify  PNC observables involving the decay of the $0^+$ level. The
experimental  arrangement (Fig.\ 7) consists of a scattering chamber with
scintillation counters to detect $\alpha$-particles emitted near $35^\circ$
and $155^\circ$. The detector geometry and the target thickness were
carefully optimized to obtain the best statistical error in the
measurement, while at the same time minimizing systematic errors.
Calculations based on the known resonance parameters predict a sharp
energy dependence of $A_z$ across the $0^+$ resonance, and an angular
dependence which changes sign between forward and back angles.  It was found
that the difference in analyzing power between back-angle and forward-angle
detectors, A(B)- A(F),  yields the largest PNC signals. However, the most
important advantage is that in the difference certain systematic errors are
reduced, because they have similar effects on A(F) and A(B). For a weak
matrix element of $-1.04$ eV\cite{ADEL84} the expected signal, taking
into account the finite spread in energy and angle, was A(B)- A(F) =
$-2.8\times 10^{-5}$. While this effect is considerably larger than for pp
scattering, the very narrow  (4 keV) low energy resonance has a cross
section with a strong dependence on energy and angle, and a large
transverse analyzing power. Therefore special methods had to be developed
to measure and control the beam energy, the beam position and the residual
transverse polarization. The targets were sputtered, 4 keV thick C enriched
in ${}^{13}{\rm C}$. Effects from target non-uniformity and $^{12}$C build-up
were
reduced by translating the target in a raster pattern during the
experiment. The beam polarization (typically $P=0.84\pm 0.01)$ was reversed
every 20 ms at the ion source. A separate measurement was made to place an
upper limit $(\Delta E < 0.45$ eV) on  the magnitude of a possible variation in
beam
energy when the proton spin is reversed  since the rapid variation of cross
section across the resonance might give a significant spurious signal. To
measure the distribution of intensity and (unwanted) transverse
polarization of the beam, 0.6 mm wide target strips were moved stepwise
through the beam in the vertical and horizontal direction. Beam position
and beam direction were controlled with a feedback system which processed
information from beam currents on slits located before and after the
target. Modulation of beam position associated with polarization reversal
was found to be $<$0.4 mm. Many spurious effects, including effects of energy
modulation,  spin misalignment, correlations between spin and beam position
and correlation between spin and  beam angle etc., were found to vary
strongly over the resonance. Fortunately, it was possible to find an energy
where most of the systematic errors nearly vanished, while the
parity-violation signal was near the maximum value. By making measurements
primarily at this particular energy, the sum of systematic errors was
reduced to $<1.5\times 10^{-6}$. The final result, A(B)-A(F) $=(0.9\pm
0.6)\times 10^{-5}$,
corresponds to a weak matrix element of $0.38\pm 0.28$ eV, {\it i.e.}, opposite
in
sign and smaller in magnitude than the theoretical expectation.

\begin{figure}
\vspace{4in}
\caption{Experimental setup for parity mixing in ${}^{14}{\rm N}$ by
measurement of $A_z$ in p+${}^{13}$C elastic scattering.  Scintillation
detectors detect scattered protons at four azimuthal angles for forward (A)
and backward (B) angles.  A NaI scintillator detects capture $\gamma$-rays.
The 4 keV thick self-supporting ${}^{13}$C target (D) is surrounded by a cold
shroud (E) to reduce buildup of contaminants.  Four-jaw adjustable slits (F
and additional sets of slits upstream and downstream of the chamber) and
steering magnets (such as G) are used to stabilize beam position and beam
direction by means of a feedback system. }
\end{figure}

\subsection{Analysis}

That high quality wavefunctions are needed for interpreting these
experiments is clear from the following argument.  Suppose one evaluates
the nuclear wavefunction in the usual $0\hbar\omega,1\hbar\omega$ shell
model basis.  Of course, a realistic wavefunction presumably includes also
an additional $2\hbar\omega$ component,
\begin{equation}
|\psi^+>=|0\hbar\omega>+\epsilon|2\hbar\omega>
\end{equation}
which is small---$\epsilon<<1$---if the simple shell model picture is valid.
Then if one calculates a typical parity conserving observable such as the
Gamow-Teller matrix element or magnetic moment, which do not connect
$|0\hbar\omega>$ and $|2\hbar\omega>$ levels, a reasonably accurate
result should obtain, since any corrections are ${\cal O}(\epsilon^2)$
\begin{equation}
<\psi^+|{\cal O}|\psi^+>=<0\hbar\omega|{\cal O}|0\hbar\omega>+
{\cal O}(\epsilon^2).
\end{equation}
However, in evaluating a parity mixing term, we are dealing with a
$1\hbar\omega$ level which can connect to either of its $0\hbar\omega$ or
$2\hbar\omega$ counterparts, whereby corrections to simple shell model
estimates
are ${\cal O}(\epsilon)$ and are much more sensitive to omission of
possible $|2\hbar\omega>$ states---
\begin{equation}
<\psi^-|{\cal H}_{\rm wk}|\psi^+>=<1\hbar\omega|{\cal H}_{\rm wk}|0\hbar\omega>
+{\cal O}(\epsilon).
\end{equation}
In fact, these
expectations are borne out, both theoretically and
experimentally.  On the theoretical side, Haxton compared
a simple $0\hbar\omega,1\hbar\omega$ evaluation with a large basis
$0\hbar\omega,1\hbar\omega,2\hbar\omega$
calculation of parity mixing between the $0^-; 1081 keV$ and $0^+; 1042 keV$
states of ${}^{18}$F, and determined\cite{HAX81}
\begin{equation}
{<0^-|{\cal H}_{\rm wk}|0^+>_{0,1,2\hbar\omega}\over <0^-|{\cal H}_{\rm wk}
|0^+>_{0,1\hbar\omega}}\simeq {1\over 3}.
\end{equation}

This calculation clearly reveals that such
${\cal O}(\epsilon)$ core polarization effects are substantial, although
clearly any such estimates are very model dependent and would seem to
offer little hope for calculational rigor.
Nevertheless, in some cases there is an opportunity to ``measure" this effect.
It was pointed out by Bennett, Lowry and Krien\cite{BEN80} and independently by
Haxton\cite{HAX81} that the form of the parity
violating nucleon-nucleon potential arising from pion exchange
\begin{equation}
V_{NN}^{PV}(\pi-{\rm exch})={i\over 2\sqrt{2}}g_{\pi NN}f_\pi
({\tau_1\times\tau_2})_3(\sigma_1+\sigma_2)\cdot\left[{{\bf p}_1-{\bf p}_2
\over 2m_N},{e^{-m_\pi r}\over r}\right]
\end{equation}
is an isotopic partner of the two-body pion exchange contribution to
the timelike component of the weak axial vector current, which is probed
in nuclear beta decay
\begin{equation}
A_0=A_0({\rm one-body})+{i\over 2}g_{\pi NN}g_A(\tau_1\times\tau_2)_\pm
(\sigma_1+\sigma_2)\cdot\left[{{\bf p}_1-{\bf p}_2\over 2m_N},{e^{-m_\pi r}
\over r}\right].
\end{equation}
Then by measuring this two body matrix element of $A_0$ in a beta transition
between levels which are isotopically related to those involved in the
weak parity mixing process, this weak pion exchange contribution to nuclear
parity violation can be calibrated experimentally.  Of course, the difficulty
with this procedure is that there is no model-independent means by which to
separate the one- and two-body contributions to $A_0$.  Nevertheless, Haxton
has pointed out that the {\it ratio}
 of such terms
\begin{equation}
{<|A_0({\rm two-body})|>\over <|A_0({\rm one-body})|>}\simeq 0.5
\end{equation}
is relatively {\it model-independent}, and by measurement of the
${}^{18}$Ne beta decay rate, one determines {\it experimentally}
\begin{equation}
{<0^-|{\cal H}_{\rm wk}|0^+>^{\rm exp}\over <0^-|{\cal H}_{\rm wk}|0^+>_
{0,1\hbar\omega}}\approx 0.35
\end{equation}
in good agreement with the full $2\hbar\omega$ theoretical estimate.
Unfortunately, such a large basis calculation is only made possible by
the feature that ${}^{18}$F is only two nucleons away from
${}^{16}$O---heavier s,d shell nuclei involve bases which are too large
for current computing capacity.

Because of the difficulties outlined above associated with extraction of
theoretical information from experimental signals arising from
nuclear experiments, physicists have tended to emphasize only p-shell and
light s-d shell nuclei for believable experiments---in particular
${}^{18}$F, ${}^{19}$F, ${}^{21}$Ne and ${}^{14}$N---and we shall discuss
each in turn.

{\bf ${}^{\bf 18}$F}:  We begin our discussion with the simplest case to
analyze---
measurement of the circular polarization in the decay of the $0^-$ 1081 keV
excited state of ${}^{18}$F to the ground state---{\it cf.} Fig. 9.

\begin{figure}
\vspace{6cm}
\caption{Energy levels for light nuclear parity violation experiments.}
\end{figure}

Because of the existence
of the $0^+$ state only 39 keV away at 1042 keV, assuming that the weak parity
mixing occurs only between these
two levels should be a good approximation.
However, the pseudoscalar 1081 keV state is an isoscalar while its scalar
1042 keV analog is an isovector.  Thus any such mixing is sensitive {\it only}
to the $\Delta I=1$ piece of the effective parity violating weak Hamiltonian
and thereby effectively only to $f_\pi$.

Another helpful feature of this case is the existence of a substantial
nuclear enhancement factor.  Because the E1 transition is between isoscalar
states this transition is isospin forbidden, leading to the comparatively
long lifetime $\tau_{1081}=27.5\pm 1.9$ps.  On the other hand the analogous
M1 transition is very fast---$\tau_{1042}=2.5\pm 0.3$fs---corresponding to
$10.3\pm 1.5$ Weisskopf units.  The resulting circular polarization then
can be written as
\begin{equation}
P_\gamma(1081)=2{\rm Re}\left[{\epsilon {\rm Amp}(M1)\over {\rm
Amp}(E1)}\right].
\end{equation}
Since for dipole emission
\begin{equation}
\Gamma_\gamma\sim |<f|{\cal O}|i>|^2\times E_\gamma^3
\end{equation}
we find
\begin{equation}
|{{\rm Amp}(M1)\over {\rm Amp}(E1)}|=({\tau_-\over \tau_+})^{1\over 2}
\left({1081\over 1042}\right)^{3\over 2}
=111\pm 8.
\end{equation}
The expected circular polarization is then
\begin{equation}
|P_\gamma(1081)|\simeq 222 {<+|{\cal H}_{\rm wk}|->\over 39 keV}
\end{equation}
and we observe that there exist two separate enhancement factors, one kinematic
and associated with the near degeneracy of the mixed states and the second
dynamic and associated with the suppression of the E1 matrix element.
  Note that because of this suppression, we quote above only the absolute
magnitude
of the circular polarization since a reliable calculation of the sign of the
electric dipole amplitude is probably out of the question.  Finally, the
isospin related
\begin{equation}
{}^{18}{\rm Ne}\rightarrow{}^{18}{\rm F}(0^-; 1081 keV)+e^++\nu_e
\end{equation}
transition can be used in order to normalize the pion exchange matrix
contribution to the weak matrix element, in the fashion described above
leading to
\begin{equation}
|P_\gamma(1081)|=4320f_\pi .
\end{equation}

In addition to the theoretical clarity of this transition it has also been
examined experimentally by five different experimental groups, all of whose
results are in agreement as shown in Table 3.  We see then that there is as
yet no evidence for the existence of a non-zero circular polarization and
that this result implies an upper bound on the value of the weak $NN\pi$
coupling which is considerably smaller than the ``best value."  While
the resulting number is certainly within the DDH bounds, it requires
considerably cancellation among the factorization, sum rule and quark model
contributions in order to achieve a result this small.

{\bf ${}^{\bf 19}$F}:  Another important result has been obtained in the
${}^{19}$F
system, where the asymmetry has been measured in the radiative decay from the
polarized $|{1\over 2}^-; 110 keV>$ first excited state down to the
$|{1\over 2}^+;g.s.>$ ground
state.  The experiment has been performed twice, and has yielded a non-zero
signal at the $10^{-4}$ level as indicated in Table 3.  Here the asymmetry
is defined via
\begin{equation}
{d\Gamma\over d\Omega}\sim 1+ A_\gamma {\bf P}_F\cdot \hat {\bf q}_\gamma
\end{equation}
and, under the assumption that only the ground and first excited state are
involved in the mixing, has the form
\begin{equation}
A_\gamma =2{<+|{\cal H}_{\rm wk}|->\over 110 keV}\times{\rm Re}{{\rm
Amp}(M1)\over
{\rm Amp}(E1)}.
\end{equation}
Here the magnetic dipole amplitude can be written in terms of the measured
(2.6289$\mu_N$) and calculated (-0.2$\mu_N$) magnetic moments of the
${1\over 2}^+$ and ${1\over 2}^-$ states respectively, while the E1 amplitude
is given in terms of the known lifetime of the ${1\over 2}^-$ level, yielding
\begin{equation}
A_\gamma={<+|{\cal H}_{\rm wk}|->\over 5.2\pm 0.4eV}.
\end{equation}
As in the case of ${}^{18}$F the pion exchange contribution to the weak
matrix element can be calibrated in terms of the measured

\begin{equation}
{}^{19}{\rm Ne}\rightarrow{}^{19}{\rm F}(0^-;110 keV)+e^++\nu_e
\end{equation}
amplitude, while the vector exchange pieces can be calculated in the shell
model, yielding
\begin{equation}
A_\gamma=-96f_\pi+35(h_\rho^0+0.56h_\omega^0).
\end{equation}
Note that since both mixed states are isodoublets the asymmetry is sensitive
to both $\Delta I=0$ and $\Delta I=1$ components of the effective weak
Hamiltonian, and we see in Tables 2,3 that the use of ``best value"
numbers yields a value for this
asymmetry which is in excellent agreement both in sign and in magnitude with
the measured number.

{\bf ${}^{\bf 21}$Ne}:  The nucleus ${}^{21}$Ne possesses states
$|{1\over 2}^+;2795 keV>$
and $|{1\over 2}^-;2789 keV>$ which are separated by only $5.74\pm 0.15$ keV.
In addition the E1 transition of the 1089 keV level down to the
${3\over 2}^+$ ground state is extraordinarily retarded,
having a lifetime $\tau =696\pm 51 ps$ and corresponding to $\sim 10^{-6}$
Weisskopf units.  One predicts then a circular polarization to be
\begin{equation}
P_\gamma(2789)=-2{<+|{\cal H}_{\rm wk}|->\over 5.74 keV}{\rm Re}\left(
{{\rm Amp}(M1)\over {\rm Amp}(E1)}\right)\times
\left({1+\delta^*_-\delta_+\over
1+|\delta_-|^2}\right)
\end{equation}
where here $\delta_-$ is the M2/E1 mixing ratio for the ${1\over 2}^-$
transition and $\delta_+$ is the E2/M1 mixing ratio for the ${1\over 2}^+$
transition.  Taking $|\delta_-|<0.6$ from experiment and $\delta_+\approx 0$
from theoretical estimates, we have
\begin{equation}
|P_\gamma|={|<+|{
\cal H}_{\rm wk}|->|\over 9.5^{+3.4}_{-0.6}eV}
\end{equation}
which indicates, as in the case of ${}^{18}$F the strong effects of both
dynamical and kinematic amplification and the fact that theory cannot
really predict the absolute sign of the highly suppressed E1 amplitude.
Since both ${1\over 2}^+$ and
${1\over 2}^-$ states are isodoublets the weak parity mixing involves
both $\Delta I=0$ and $\Delta I=1$ components of the weak Hamiltonian and
a shell model calculation gives
\begin{equation}
P_\gamma=29500f_\pi+11800(h_\rho^0+0.56h_\omega^0).
\end{equation}
Comparing with the analogous calculation for the case of ${}^{19}$F we see
that the vector- and pion-exchange amplitudes come in with opposite signs,
indicating the difference the ``odd-proton" (${}^{19}$F) and ``odd-neutron"
(${}^{21}$Ne) nuclei and that the nuclear enhancement factors are nearly
a factor of 300 larger in the case of ${}^{21}$Ne due to the near degeneracy
and
strong suppression of the E1 decay amplitude discussed above.  Using the
``best value" numbers one finds that sizable cancellation between the
pion- and vector-exchange components takes place so that the predicted
and experimental size
for the circular polarization are in agreement, but this requires a
significant value for $f_\pi$ which is inconsistent with the upper
bound determined from ${^{18}}$F.  We shall have more to say on this
problem in a later section.

{\bf ${}^{\bf 14}$N}: The only p-shell nucleus to make our list is ${}^{14}$N
for which there exist states $|0^-,8776$ keV> and $|0^-,8624$ keV> which are
separated by only 152 keV.  Both states are isotopic triplets but calculation
indicates that mixing is due predominantly to the $\Delta I=0$ component
of the effective weak interaction.  In this case one observes $A_z$ for the
delayed proton emission from the $0^+$ state and there
exists a dynamical enhancement factor of $[\Gamma_p(0^-)/\Gamma_p(0^+)]^
{1\over 2}\approx 11$.  While at first glance, one might believe that the
shell model analysis might be relatively reliable, inasmuch as a p-shell
nucleus is involved, the problem is that the natural parity state in
${}^{14}$N is predominantly $2\hbar\omega$ in character unlike previously
studied parity doublets wherein the natural parity state is primarily
$0\hbar\omega$.  Thus a very large shell model basis is required and
various approaches lead to predictions
\begin{equation}
-1.39 eV\leq <-|{\cal H}_{\rm wk}|+> \leq -0.29 eV
\end{equation}
if the DDH value of $h_\rho^{(0)}$ is employed.  The discrepancy with the
positive sign of the measured number is disturbing and remains to be
explained.

Notice that we have not attempted to analyze the ${}^{16}$O($2^-)$ alpha
decay problem.  That is because mixing occurs with any of the many
$2^+$ levels of ${}^{16}$O and there is no reason to favor any particular
level.  Thus the calculation, while it has been performed, is thought to
be rather uncertain, even though achieving a result in agreement with the
experimental number.\cite{BRS80}

\section{Nuclear Parity Violation and Statistical Methods}

Above we spoke despairingly about the use of heavy nuclei in experiments
involving
nuclear parity violation because of the lack of believable nuclear
wavefunctions.
Recently, however, it has become clear that in some cases one {\it can}
actually
employ
heavy nuclei by exploiting their statistical properties.  The case in point
involves a set of high precision longitudinally polarized epithermal neutron
scattering
measurements performed on a series of heavy nuclei at LANSCE by Frankle
{\it et al.}\cite{FRAN91,YUA91}
The first round of such experiments involved ${}^{239}$U and ${}^{232}$Th
targets, and a set of transmission experiments revealed parity-violating
asymmetries, with a statistical
significance of greater than 2.5 standard deviations,
in three U states and seven Th states.  Now when such an
epithermal neutron is captured, the resulting compound nuclear state is
made up of linear combinations of $10^5$ to $10^6$ single-particle
configurations
so that one would expect that a statistical model of the nucleus, with
observables treated as random variables should be quite sufficient.  In such
a picture one expects to find occasional p-states in a large background of
s-wave resonances, and the roughly one third of these which are ${\rm
p}_{1\over 2}$
character can mix with both nearby and distant ${\rm s}_{1\over 2}$ levels,
leading
to the observed parity violating asymmetries.  The mixing matrix elements of
the
weak interaction should be of single particle character.  The experimenters
interpret the measured longitudinal asymmetries for compound nuclear
states in the region $  10 eV<\Delta E<300 eV$ in terms of a mean
squared matrix element $M^2=<|<\mu|{\cal H}_{\rm wk}|\nu>|^2>_{\Delta E}$.
Then, using the ergodic theorem, this number can be identified with the
ensemble
average, yielding the result
\begin{equation}
M=[{\rm Av}\{|<\mu|{\cal H}_{\rm wk}|\nu>|^2\}]^{1\over
2}=0.58^{+0.50}_{-0.25}\ meV.
\end{equation}
The size of the mixing is about what one might expect, as the density of states
in this region is about a thousand times larger than found in light nuclei,
changing the typical $1\ eV$ value found for typical isolated weak levels
to the $1\ meV$ determined above.  However, it is possible to be somewhat more
quantitative by using the microscopic framework developed by French, who
relates experimental and theoretical mean square matrix elements in terms
of a strength $\alpha$ of a schematic symmetry violating interaction
$\alpha U_2$ where $U_2$ is the residual shell-model interaction acting
in a model space.  With the value $<|<\mu|U_2|\nu>|^2>_{\Delta E}=2.6\ keV^2$
for ${}^{239}$U from ref. one finds then $\alpha_P^2=$.  Using the G-matrix
formalism and the closure approximation Johnson {\it et al.} have attempted to
make contact between the statistical formalism of French and the underlying
weak Hamiltonian developed in ref.\cite{JOHN91}  Their results are summarized
in
Table 4 for three different sets of weak interaction couplings, as given
by the DDH best values, improved best values as later calculated by
Feldman {\it et al.},\cite{FELD91} and empirical numbers generated by
Adelberger and Haxton\cite{ADEL85}.
As can be seen, all are in agreement with the experimental number,
indicating that the overall scale of the parity-violating
interaction is basically correct.

\begin{table}
\caption{Experimental and theoretical values for weak mixing parameters
as determined in epithermal neutron scattering.}
\begin{center}
\begin{tabular}{|c|c|c|}\hline
Interaction &$\alpha/ G_Fm_\pi^2$ & M(meV) \\
\hline
DDH&2.67&0.98\\
ref. 8&1.54&0.52\\
ref. 19b&1.07&0.39\\
\hline
\end{tabular}
\end{center}
\end{table}

In the first data taken by this group it was found that out of seven
levels in ${}^{232}$Th exhibiting parity violation all seven had the same
{\it sign} for the asymmetry!  This result appeared to be in strong
contradiction with the presumed statistical nature of the mixing process,
which would seem to require roughly equal positive and negative values.
However, with the taking of additional data on other nuclei
the number of data points on either side of zero has evened out somewhat, and
at the present time the thorium result is thought to be due to some quirk of
nuclear structure.

\section{A New Probe of Nuclear Parity Violation: the Anapole Moment}

A somewhat different approach to the problem of measuring NNM matrix elements
was recently proposed in the realm of electron scattering.  The idea here is
somewhat subtle and so requires a bit of explanation.  Suppose that one is
considering the most general matrix element of the electromagnetic current
between a pair of nucleons.  The most general form allowed by spin and gauge
invariance considerations is
\begin{eqnarray}
<N({\bf p}')|V_\mu^{\rm em}(0)|N({\bf p})>=\bar{u}({\bf p}')[f_1(q^2)\gamma_\mu
-i{f_2(q^2)\over 2m_N}\sigma_{\mu\nu}q^\nu\nonumber\\
+{f_A(q^2)\over m_N^2}(q^2\gamma_\mu -q_\mu\gamma^\nu q_\nu)\gamma_5
-i{f_E(q^2)\over 2m_N}\sigma_{\mu\nu}q^\nu\gamma_5]u({\bf p})\nonumber\\
\quad
\end{eqnarray}
where $q=p-p'$ is the four-momentum transfer.  Here $f_1(q^2),f_2(q^2)$ are
the familiar charge, magnetic couplings respectively.  The
remaining two terms involving $f_A(q^2), f_E(q^2)$ may look unfamiliar as
they are usually omitted on the grounds of parity conservation.  However,
if one allows for the possibility that parity is violated, then such terms must
be included.  The term involving $f_E(q^2)$ is found to be time reversal
violating as well as parity violating and corresponds to a nucleon electric
dipole moment.  On this basis, we can safely omit it.  However, the term
$f_A(q^2)$ is time-reversal allowed and must be retained in a general
analysis.  It is generally called the ``anapole moment" and would appear to be
a
fundamental property of the nucleon.  However, this is {\it not} the case.
In fact the anapole moment is not strictly speaking
an observable since estimates of its
size depend upon the weak gauge in which one chooses perform the calculation.
How can this be?  The resolution of the paradox lies in the way such a
quantity could be measured---{\it i.e.} via parity violation in electron
scattering.
In such an experiment the total parity violating signal arises due
to the coherent sum of photon exchange diagrams involving the anapole moment
plus diagrams involving both photon and $Z^0$ boson exchange, as shown in
Fig.\ 9.  Of course, the {\it sum} of these effects must be an observable and
independent of gauge.  However, it is not required that each diagram
individually
be gauge independent.  (In this way the anapole moment is like the neutrino
charge radius, which is similarly gauge-dependent.)

\begin{figure}
\vspace{6cm}
\caption{Parity violating electron scattering via anapole and direct
Z-exchange mechanisms.}
\end{figure}

It would seem then that attempts to measure the anapole moment cannot possibly
be meaningful.  However, this is fortunately not the case.  In order to see
how this comes about we divide the anapole moment into its one-body and
many-body components.   For the one-body (impulse-approximation) term
we find
\begin{equation}
{\bf V}=-{1\over m_N^2}\sum
f_A^i[\sigma_i\nabla^2-\sigma_i\cdot\nabla\nabla]r^2
\end{equation}
where $f_A^i$ is the anapole moment of the ith nucleon.  The magnitude of this
term
then is determined by the properties of the nucleon and not the nucleus and, of
course, its size is gauge dependent.  On the other hand, many body
contributions
such as those generated by single meson exchange within the nucleus, as shown
in Fig.\ 11, are gauge independent and grow as the square of the {\it nuclear}
radius---$f_A^{\rm many-body}\sim <R^2>\sim A^{2\over 3}$.  In the limit
$A\rightarrow\infty$ then this many body and gauge independent quantity must
dominate over its gauge dependent single body counterpart and the anapole
moment will be an observable.\cite{MUS91}  In fact, calculations have shown
that
in the real world many cases exist for even moderately heavy nuclei where the
many
body component should be the dominant effect.  This occurs both in the case
of heavy nuclei such as ${}^{133}$Cs where the pion exchange contribution has
been estimated to be a factor of three larger than the tree-level
$Z^0$-exchange
piece and even in ${}^{19}$F, where the existence of nearby
${1\over  2}^+,{1\over 2}^-$ levels enhances the many body component by a
factor of two and makes it comparable to the tree-level piece.\cite{HAX89}  In
these cases
or others then, to the extent that the many body term could be measured and
that its size is dominated by the diagrams shown in Fig.\ 11, this would
provide in principle an independent way of measuring the weak parity violating
NNM
couplings.

\begin{figure}
\vspace{7cm}
\caption{Single meson exchange diagram within the nucleus.}
\end{figure}

At the present time, there is some indication that an anapole
moment has been seen via a study of parity nonconserving signals from
different hyperfine levels in atomic Cs(ref.\cite{NOE88}).  The measured signal
is of
the same order but somewhat larger than theoretical
expectations.\cite{PAF88}
However, this is only preliminary and it will
be some time before it will be known if this technique represents a viable
approach to the study of nuclear parity violation.

\section{How Large are the Weak Couplings}

In our analysis above we have consistently compared the experimental results
with theoretical predictions based on the ``best value" guesses of DDH for
the weak NNM vertex functions.  However, it is also possible and desirable
to determine such couplings purely empirically.  If the couplings obtained
in this fashion are found to be mutually consistent they then form a
benchmark against which past and future particle physics calculations can
be calibrated.  Of course, there are many parameters involved and many
parity violating experiments so a simple statistical fit is probably
not appropriate.  However, a little thought reveals that the process can in
principle be made meaningful, as pointed out by Haxton and Adelberger.
On the experimental side, the data set to be fit was restricted to those
cases wherein one has both good statistical precision as well as
a reasonable expectation for a reliable theoretical calculation.  This
limits things to the $\vec{\rm p}{\rm p},\vec{\rm p}\alpha,{}^{18}{\rm F},
{}^{19}{\rm F},$ and ${}^{21}{\rm Ne}$ systems.  On the theoretical side, a few
prejudices from the DDH analysis were employed in order to characterize
all results in terms of just two free parameters---$f_\pi$ and $\eta$, which
characterizes the SU(6)
breaking in the calculation and interpolates between factorization ($\eta
=0)$ and pure SU(6)$_W$ ($\eta=1$) results.  The result of this fit
is shown in Table 5 and in a different form in Fig.\ 12.

\begin{figure}
\vspace{6cm}
\caption{Experimental constraints on weak couplings.}
\end{figure}

As can be seen
therein, there exists a fundamental problem in that while the ${}^{18}$F
data require a very small value for $f_\pi$, a much larger value is needed
in order to cancel against $h_\rho^0$ to produce the very small circular
polarization seen in the decay of ${}^{21}$Ne.  In fact, were ${}^{21}$Ne
to be omitted as a constraint a very satisfactory fit of the remaining
experiments would result as shown in Table ???.

\begin{table}
\caption{Fitted values for weak
NNM couplings.  All numbers are to be
multiplied by the factor $3.8\times 10^{-8}$.}
\begin{center}
\begin{tabular}{|c|c|c|}\hline
 &Range&Fitted value\\
\hline
$f_\pi$&$0\rightarrow 30$ & 6\\
$h_\rho^0$&$30\rightarrow -81$& -15\\
$h_\rho^1$&$-1\rightarrow 0$& -0.5\\
$h_\rho^2$&$-20\rightarrow -29$& -20\\
$h_\omega^0$&$15\rightarrow -27$& -13\\
$h_\omega^1$&$-5\rightarrow -2$& -1.5\\
\hline
\end{tabular}
\end{center}
\end{table}

So what is the problem?  Of course, one possibility is that the
simple and appealing single meson exchange picture developed above is
not appropriate.  However, in view of the success obtained with the
corresponding meson-exchange approach to the ordinary nucleon-nucleon
potential this seems unlikely.  Rather it would seem that the most
likely explanation lies in our inability to perform an adequate large
basis calculation for nuclear systems.  Indeed the inclusion
of core polarization effects for the lighter ${}^{19}$F and ${}^{19}$F
systems has already been shown to lead to very substantial changes.
Likewise recent calculations by Horoi and Brown have indicated the importance
of inclusion of $3\hbar\omega$ and $4\hbar\omega$ states in the shell model
basis for calculations involving parity violation in p- and s,d-shell
nuclei.\cite{HOR94}

\section{The Future of Nuclear Parity Violation}

Above we have examined the many attempts to understand the phenomenon of
nuclear
parity violation from the first measurements during the 1950's until the
present day.  We have seen that despite the many and elegant experiments which
have been completed and the extensive theoretical effort which has gone into
this problem, many difficulties still remain and it is not yet clear that
the simple meson exchange picture is able to explain all the varied results.
Were this to be verified it would be very surprising since a similar single
meson exchange picture {\it is} remarkably successful in explaining all aspects
of
the ordinary nucleon-nucleon potential.  Nevertheless it remains to be seen.
In the mean time, it is interesting to ask whether say by the end of the decade
experiments will be
available to aid in this process and/or whether new theoretical work will be
able to add new illumination on the mechanism of nuclear parity violation.

In the case of theoretical work, we are somewhat pessimistic.  Barring some
clear breakthrough it is unlikely that things will change much during this
period.  One might
think that lattice methods might be of help here, but when one is dealing
with three-hadron matrix elements of a four-quark operator, we are still
far from being able to make reliable calculations.  In the case of non-lattice
procedures, the only semi-rigorous technique which has yet to be applied
consistently to this problem is that of QCD sum rules.  However, again
the complex hadron states and four-quark operators make this a severe
challenge.  One area which deserves further exploration is the role of
strangeness.  Recent experiments involving the spin structure of nucleons
and neutrino-nucleon
scattering have hinted that the nucleon may have a
significant strange quark component, which has been neglected in previous
calculations of weak NNM couplings.  Combined with a new calculation which
includes strange quark contributions to the effective weak Hamiltonian,
one might anticipate a few changes, especially in the pion emission
amplitudes.  However, such calculations are very difficult and at the
present time are very speculative.

In the case of experiment, we are more fortunate, with a number
of possible new results coming on line within the next couple of years.  One
which has
been in the planning stage for many years is a TRIUMF measurement of the
asymmetry in longitudinally polarized pp scattering at 240 MeV.  The
significance
of this energy is that according to known phase shifts this is where any
effect due to S- and P-wave mixing and thereby a roughly equal contribution
from rho and omega exchange effects cancels out, leaving sensitivity
primarily to rho exchange contributions in P-D wave interference.

Another arena where it is possible that a new experiment could make a major
impact is the measurement of parity mixing between the ground state and
first excited state of ${}^{19}$Ne, which are the isotopic analogs of the
mixed states which are studied in the ${}^{19}$F experiments.  Since both
states are isotopic doublets, only the $\Delta I=0,1$ components of
${\cal H}_{\rm wk}$ are operative, and by combining the results of the
Ne and F measurements an unambiguous separation of the $\Delta I=0$ and $\Delta
I=1$ components would result.  This could enable an additional
and welcome confirmation of the calibration of the pion exchange component,
as well as an independent measurement of the size of $f_\pi$.

As discussed above, we expect that continued parity violating electron
scattering experiments during the next few years will lead finally to
a measurement of the anapole moment and that theoretical
work may enable extraction of the NNM couplings in this unique fashion.

Finally, neutron scattering measurements will continue both at Los Alamos,
where studies of heavy nuclei have already indicated the power of statistical
methods, as well as at NIST where the successful Grenoble program will
be extended to lighter nuclei, which are hopefully amenable to clearer
theoretical interpretation.

In summary then in both theoretical and experimental areas one sees
the need for
additional and improved work and we set the challenge that perhaps by the
millenium one may finally put this problem to rest.

\end{document}